# CCR 2.0: Refinement, Unary, and Relational Separation Logic

YOUNGJU SONG, MPI-SWS, Germany
MINKI CHO, Seoul National University, Korea

In recent years, great progress has been made in the field of formal verification for low-level systems. Many of them are based on one of two popular approaches: *refinement* or *unary separation logic*. These two approaches are very different in nature and offer complementary benefits in compositionality.

Recently, to fuse these benefits into a single unified mechanism, a new approach called Conditional Contextual Refinement (CCR 1.0 for short) was proposed. In this paper, we advance CCR 1.0 and provide novel and intuitive reasoning principles, resulting in CCR 2.0. Achieving this goal was challenging due to non-trivial counterexamples which necessitated elegant changes to the model of CCR 1.0. On top of CCR 2.0, we show how to *fuse* the benefits of refinement, unary separation logic, and *also* relational separation logic.

Our results are formalized in Rocq.

## 1 INTRODUCTION

Recent years have seen significant advances in formal verification of low-level systems [Appel 2014; Chajed et al. 2019; Gu et al. 2016; Klein et al. 2009]. These verification techniques primarily employ either *refinement* or *separation logic*—with separation logic further categorized into *unary* and *relational* variants. Each approach offers distinct compositional properties and adequacy guarantees.

Recently, to fuse these benefits in a unified mechanism, a new approach called Conditional Contextual Refinement (CCR 1.0 for short) [Song et al. 2023] was proposed. Song et al. [2023] demonstrated CCR 1.0's distinctive advantages over all three existing approaches, showing great promise. Specifically, their verification *fuses* the benefits of unary separation logic and refinement, enabling a fundamentally new verification strategy.

However, benefits of CCR 1.0 (which fuses *unary* separation logic and refinement) and *relational* separation logic are still kept separate—roughly speaking, CCR 1.0 allows establishing Hoare triple (of *unary* separation logic) via vertical composition (of refinement), which is inherently different from Hoare quadruple (of *relational* separation logic). A question that is left unanswered is:

Can we fuse the benefits of *all three* (unary, relational separation logic and refinement) styles?

In this paper, we present CCR 2.0, which answers this question with "yes". This fusion enables new verification strategies that were not possible in any existing framework (including relational separation logics and CCR 1.0).

To see more clearly what this "fusion" means and what becomes newly possible, see Fig. 1. The table summarizes existing related work and how CCR is positioned among them (we use CCR to refer to both CCR 1.0 and CCR 2.0). While we include only a few representative examples, the first row represents refinement frameworks generally, the second row represents unary separation logics, and the fourth and fifth rows represent traditional and modern relational separation logics.

Refinement Calculus verifies correctness in the form of contextual refinements, and its key strength is vertical compositionality (column **VC**)—$T \sqsubseteq M \wedge M \sqsubseteq S \implies T \sqsubseteq S$. Iris program logic and VST support ownership-based reasoning for shared states (column **USL**).

Relational separation logics use separation logic to establish contextual refinements. Specifically, it uses a Hoare Quadruple $\{P\}\, T \leq S\, \{Q\}$, which allows relational reasoning (column **RSL**): one suffices to prove relatedness of two sides, rather than proving the safety (which could be false from the first place). The adequacy theorem of Hoare Quadruple says that, when $P$ and $Q$ are certain

Authors' addresses: Youngju Song, MPI-SWS, Germany, youngju@mpi-sws.org; Minki Cho, Seoul National University, Korea, minki.cho@sf.snu.ac.kr.

| | VC | USL | RSL | VC × USL | VC × USL × RSL | SL Interface |
|---|---|---|---|---|---|---|
| Refinement Calculus | ● | ○ | ○ | N/A | N/A | N/A |
| Iris Program Logic, VST | ○ | ● | ○ | N/A | N/A | ● |
| CCR 1.0 (**Section 2.1**) | ● | ● | ○ | ◐ | N/A | ◐ |
| Rel. SL (Yang [2007]) | ● | ○ | ● | N/A | N/A | ● |
| Simuliris, DimSum | ● | ● | ● | ○ | ○ | ● |
| CCR 2.0 (**Section 2.2**) | ● | ● | ● | ● | ● | ● |

Fig. 1. Comparison of related work (VC = vertical compositionality; SL = separation logic; USL = Unary SL; RSL = Relational SL).

trivial conditions, the Quadruple can be turned into a refinement judgment $T \sqsubseteq S$. Thus, one can enjoy vertical compositionality after applying the adequacy theorem (column **VC**).

Moreover, modern relational separation logics like Simuliris and DimSum support unary separation logic reasoning through "focus" rules, allowing one side of the program to be analyzed and executed independently while the other side remains fixed (column **USL**).

In the next section, we will show how CCR is positioned against these existing works with a running example. To streamline the presentation, we were compelled to organize it into two stages:

In (**Section 2.1**), we show how CCR 1.0 *fuses* unary separation logic and refinement. CCR 1.0 is not a mere disjoint union of two techniques: its fusion enables a fundamentally new feature that utilizes vertical compositionality and separation logic at the same time.

The column **VC** × **USL** denotes any form of logical rule that allows vertical composition of refinements together with unary separation logic conditions on it. For instance, consider the following vertical composition rule for some Hoare Quadruples and some composition operator $\ltimes$:

$$\frac{\{\mathsf{P}_0\}\ b_t \le b_m\ \{\mathsf{Q}_0\} \qquad \{\mathsf{P}_1\}\ b_m \le b_s\ \{\mathsf{Q}_1\}}{\{\mathsf{P}_0 \ltimes \mathsf{P}_1\}\ b_t \le b_s\ \{\mathsf{Q}_0 \ltimes \mathsf{Q}_1\}} \quad \text{(HQ-VCOMP)}$$

saying that if $b_t$ refines $b_m$ (with conditions $\mathsf{P}_0, \mathsf{Q}_0$) and $b_m$ refines $b_s$ (with conditions $\mathsf{P}_1, \mathsf{Q}_1$), those two can be vertically composed to show that $b_t$ refines $b_s$ (with conditions somehow composed).

While looking sensible, no existing relational separation logic supports this form of compositionality. To the best of our knowledge, CCR 1.0 is the only work that ever presented a composition akin to the above rule, although it only supported *unary* separation logic conditions and not *relational* separation logic conditions in their Hoare-Quadruple-like judgment (thus ○ in column **VC** × **USL** × **RSL**). Also, while CCR 1.0 demonstrated this composition with examples, they did not provide a general logical rule as above. As it turns out, the underlying model of CCR 1.0 does not admit a general composition rule as above for arbitrary unary separation logic conditions, and this fundamentally limits proof reuse (thus ◐ in column **VC** × **USL**).

In (**Section 2.2**), we show how CCR 2.0 *fuses* relational separation logic and CCR 1.0. Just like before, CCR 2.0 is not a mere disjoint union of two different techniques: its fusion enables a fundamentally new verification. CCR 2.0 adds support for relational separation logic predicates, and that support is naturally extended to **VC** × **USL** to the above composition theorem. As a result, CCR 2.0 allows unary separation logic predicates to specify arbitrary conditions on the external states, and also allows use of relational separation logic predicates in its verification (thus ● in column **VC** × **USL** × **RSL**), and this newly enables verification of refinements that were fundamentally unsupported in CCR 1.0.

In (**Section 4**), we discuss two fundamental improvements we made in CCR 1.0 (resulting in CCR 2.0). It gives a general logical rule for the above composition (thus ● in column **VC** × **USL**) and also offers an interface that hides all the model-level details of separation logic and provides higher-level logical rules to the user (column **SL Interface**). In fact, these are the most novel and challenging part of this work which improves very core foundation of CCR in a non-trivial way: Adding support for **RSL** can be seen as a flagship example that is enabled by this update. The relational logic we have implemented is equivalent to the relational logic in DimSum (or, Simuliris without concurrency), but implementing it on top of CCR 2.0 opens up a new capability (column **VC** × **USL** × **RSL**) that is not possible in DimSum (and Simuliris).

For expository purposes, we defer the discussion on improving core foundation of CCR to **Section 4** and afterwards, while focusing on new capabilities in CCR 2.0 in early sections.

Finally, we have omitted rather less important or orthogonal features from Fig. 1, including termination sensitivity and concurrency. The correctness result of CCR (based on coinductive techniques) is termination sensitive, but does not support step-indexed techniques. CCR do not support concurrency, and adding support for concurrency is a separate ongoing project.

**The results in this paper are fully formalized in the Rocq proof assistant [Song and Cho 2025].**

## 2 CCR 2.0 BY EXAMPLE

### 2.1 Unary Separation Logic, Refinements, and Fusing Both in CCR

Let us begin with a brief overview of existing approaches, and see what CCR brings to the table. Specifically, we discuss the differences between these approaches in terms of (1) the end result (adequacy), and (2) a compositional notion of correctness that is used to establish the end result.

***Unary separation logic.*** In unary separation logic [Appel 2014; Jung et al. 2015; Ley-Wild and Nanevski 2013; Pym et al. 2004], the end result is (usually) the safety (*i.e.*, absence of undefined behavior) of a given program.

The compositional notion of correctness used here is a Hoare triple. Using Hoare triples, one can verify a component (a function or a code block) at a time, instead of verifying the whole program in a monolithic way. Specifically, the triple $\{P\}\ \mathtt{C}\ \{Q\}$ (we use purple color for separation logic conditions) means that given a *precondition* $P$, a program component, C is safe to execute, and if it terminates, the resulting state will satisfy the *postcondition*, $Q$. These *conditions* are at the heart of the Hoare triple: thanks to these, each component can be modularly verified by utilizing just the conditions ($P, Q$) without looking into the full implementation details (C) of other components.

In separation logic, the conditions are "spatial" propositions, an enhancement from "pure" propositions. These spatial propositions can specify interesting conditions on shared states, regardless of whether such states are apparent to the code being verified (*e.g.*, a spin lock implementation does *not* have access to protected shared states, but its specification can state ownership over those states). The notion of ownership gives rise to the famous "frame rule", which allows modular and compositional verification for programs with shared states.

***Refinement.*** In the refinement approach [Back and Wright 2012; Choi et al. 2017; Dijkstra 1972; Kragl and Qadeer 2018; Kragl et al. 2020; Liang and Feng 2016; Wirth 1971], the end result is a refinement between an implementation written in a programming language like C, and a mathematical abstraction: $T \sqsubseteq S$ ($T$ for the target, the more concrete side, and $S$ for the source, the more abstract side). This result is often more useful than mere safety: specifications of interactive systems such as operating systems or hypervisor are very naturally given in the form of refinement [Li et al. 2021].

The compositional notion of correctness here is, again, the refinement. Specifically, the refinement comes with a notion of *vertical compositionality*. This allows one to prove multiple step-wise refinements ($T \sqsubseteq M_1$, $M_1 \sqsubseteq M_2$, …, $M_n \sqsubseteq S$), and *vertically compose* them to establish the end result ($T \sqsubseteq S$). This form of verification is often called *gradual*, *layered*, or *step-wise* verification.

$T_{\text{Map}}$
```
private data: *long := NULL

def init(sz: long) {
  data := calloc (sz * 8)
}
def get(k: long): long {
  return * (data + k)
}
def set(k: long, v: long) {
  print (k, v);
  * (data + k) := v
}
```

$M_{\text{Map}}$
```
private map: long → long := λk. 0
private size: Option long := None

def init(sz) {
  assert(size = None);
  size := Some(sz) }
def get(k) {
  assume(0 ≤ k < size?);
  return map(k) }
def set(k, v) {
  print (k, v);
  assume(0 ≤ k < size?);
  map[k] := v }
```

$S_{\text{Map}}$
```
private map := λk. 0

def init(sz) {
  skip
}
def get(k) {
  return map(k)
}
def set(k, v) {
  print (k, v);
  map[k] := v
}
```

$C_{\text{Map}}$

$\forall sz.\ \{\underline{uninit}\}$ init(sz) $\{\top\}$
$\forall k.\ \{\top\}$ get(k) $\{\top\}$
$\forall k\, v.\ \{\top\}$ set(k,v) $\{\top\}$

$E_{\text{Map}}$

$\forall sz.\ \{\underline{uninit}\}$ init(sz) $\{\ast_{k\in[0,sz)}\ k \mapsto_{\text{Map}} 0\}$
$\forall k\, v.\ \{k \mapsto_{\text{Map}} v\}$ get(k) $\{r.\ r = v \wedge k \mapsto_{\text{Map}} v\}$
$\forall k\, w\, v.\ \{k \mapsto_{\text{Map}} w\}$ set(k,v) $\{k \mapsto_{\text{Map}} v\}$

Fig. 2. An implementation $T_{\text{Map}}$, its intermediate abstraction $M_{\text{Map}}$, and its final abstraction $S_{\text{Map}}$ (above). Conditions used in the lower refinement $C_{\text{Map}}$, and conditions used in the upper refinement $E_{\text{Map}}$ (below).

Oftentimes, the notion of refinement—especially the well-known *contextual refinement* and its variants—comes with an additional notion of composition, called *horizontal compositionality*.

$$(T_1 \sqsubseteq S_1 \quad \wedge \quad T_2 \sqsubseteq S_2) \quad \Longrightarrow \quad T_1 \oplus T_2 \sqsubseteq S_1 \oplus S_2$$

where $\oplus$ is a module linking operator. This form of compositionality allows one to verify each component, 1 and 2 here, separately.

However, this does not offer as powerful a decomposition mechanism as in separation logic. This is because in establishing these “small” refinements (*e.g.*, $T_1 \sqsubseteq S_1$), the kind of specifications (conditions) one can assume for each function is not as expressive as separation logic. In basic contextual refinement, one can assume nothing, and although there are more sophisticated versions that allow simple conditions based on types [Frumin et al. 2021] or *pure* pre/postconditions [Back and Wright 2012], none of them supports general *separation logic* pre/postconditions.

Henceforth, we use the term *un*-conditional refinements or *vanilla* refinements to refer to existing notions of refinements to distinguish them from conditional refinements.

***A motivating example.*** CCR fuses the benefits of these two approaches: it enjoys the benefit of refinement (layered verification) and the benefit of separation logic (highly flexible component-wise verification enabled by expressive, function-wise separation logic specifications) at the same time.

To see this more clearly, let us take a look at Fig. 2. [1] The implementation, $T_{\text{Map}}$, is a simple key-value storage and comprises three methods: `init`, `get`, and `set`. To use this module, one first needs to invoke `init` which zero-initializes the storage with the given size, `sz`. After that, `get` and `set` could be invoked which retrieves and stores the value corresponding to the given key. `set` additionally “prints” given arguments, which is a visible event. Here, the module-private (or, module-local) variable `data` is accessible internally but is completely hidden from the outside.

[1] Borrowed and modified from Song et al. [2023] under authors’ permission.

The abstraction, $S_{\mathsf{Map}}$, abstracts away the low-level details—memory accesses—and instead gives an abstract view of the computation. Specifically, its module-private variable `map` stores a mathematical function of type `long` → `long` with zero as initial values. Then, the method `get` retrieves and `set` stores the value from this `map` variable instead of memory.

Following the common style in refinement-based verifications, we want to verify the final refinement $T_{\mathsf{Map}} \sqsubseteq S_{\mathsf{Map}}$ in a step-wise manner, *i.e.*, by composing $T_{\mathsf{Map}} \sqsubseteq M_{\mathsf{Map}}$ and $M_{\mathsf{Map}} \sqsubseteq S_{\mathsf{Map}}$ ($M_{\mathsf{Map}}$ will be explained soon). However, existing approaches fall short in verifying this.

First, let us begin with refinements. Its upside is that we can straightforwardly state the stepwise verification as above $T_{\mathsf{Map}} \sqsubseteq M_{\mathsf{Map}}$ and $M_{\mathsf{Map}} \sqsubseteq S_{\mathsf{Map}}$. Its downside is that the refinement $T_{\mathsf{Map}} \sqsubseteq S_{\mathsf{Map}}$ does *not* hold in the first place. To be concrete, consider the following ill-formed call sequence that calls `init` twice: `init(1); set(0,42); init(1);` *print*`(get(0))`. In $T_{\mathsf{Map}}$, it will print `0`, whereas in $S_{\mathsf{Map}}$, it will print 42.

Next, let us consider unary separation logic. Its upside is that one could use the expressive language of separation logic to specify the intended protocol between `Map` module and the outside world, shown in $E_{\mathsf{Map}}$. Here, *uninit* is a globally unique and non-fungible token that gets consumed when calling `init`. Requiring this token in the precondition of `init` rightfully enforces the client to call `init` at most once. The *points-to* connective, $\mathsf{k} \mapsto_{\mathsf{Map}} \mathsf{v}$, denotes that the key `k` currently contains the value `v`, and also the ownership to call `get` and `set` with that key. With these points-to, clients can be assured that the return value of `get` coincides with their expectation. However, the downside of separation logic is that it lacks the notion of stepwise verification as in refinements.

***Fusing both in* CCR.** CCR enhances vanilla refinement with separation logic conditions:

$$C \vDash T \sqsubseteq S$$

This notion called *conditional* refinements, is now equipped with $C$, a partial map from function names to their specifications (*i.e.*, pre/postconditions). With this, we will establish the end result of $E_{\mathsf{Map}} \vDash T_{\mathsf{Map}} \sqsubseteq S_{\mathsf{Map}}$. Intuitively, this refinement judgment additionally dictates that each function will satisfy its pre/postconditions. To be specific, when proving $E_{\mathsf{Map}} \vDash T_{\mathsf{Map}} \sqsubseteq S_{\mathsf{Map}}$, we can *rely on* the precondition of `init` at the beginning of the function and should *guarantee* the postcondition before it returns. On the other hand, when proving the client of `Map`, say Clnt, the conditional refinement judgment $E_{\mathsf{Map}} \vDash T_{\mathrm{Clnt}} \sqsubseteq S_{\mathrm{Clnt}}$ will enforce one to *guarantee* the precondition of `init` at the call site, and allow one to *rely on* postcondition when it returns.

With conditional refinements, the proposition $E_{\mathsf{Map}} \vDash T_{\mathsf{Map}} \sqsubseteq S_{\mathsf{Map}}$ is now (at least) true, thanks to the conditions $E_{\mathsf{Map}}$ ensuring that `init` is called at most once.

Now, we can decompose our end goal into the following two stepwise refinements on the premise:

$$C_{\mathsf{Map}} \vDash T_{\mathsf{Map}} \sqsubseteq M_{\mathsf{Map}} \quad \wedge \quad D_{\mathsf{Map}} \vDash M_{\mathsf{Map}} \sqsubseteq S_{\mathsf{Map}} \quad \Longrightarrow \quad E_{\mathsf{Map}} \vDash T_{\mathsf{Map}} \sqsubseteq S_{\mathsf{Map}}$$

Here, we carefully choose $C_{\mathsf{Map}}$ so that the lower (the left on the premise) refinement only concerns *abstraction* of memory (`data` into `map`) and the upper (the right on the premise) refinement only concerns giving useful separation logic specifications $D_{\mathsf{Map}}$ to the client. Here, $D_{\mathsf{Map}}$ is equal to $E_{\mathsf{Map}}$ except that it does not have *uninit* in the precondition of `init`.

Then, CCR 2.0's compositionality theorem (column **VC** × **USL**) composes these two conditional refinements into the one on the right. This is in contrast to the vanilla transitivity (column **VC**), which lacks separation logic conditions. Compared to CCR 1.0, CCR 1.0 also provides roughly the same proof obligation for this specific example, but their underlying model of conditional refinement does not admit a general composition theorem (thus ◐ in column **VC** × **USL**). This fundamentally limits proof reuse when one has multiple implementations with a same specification (more discussion on §4.1). Finally, no relational separation logic supports any form of vertical composition together with separation logic conditions like above (thus ○ in column **VC** × **USL**).

Let us take a look at the lower refinement $C_{\mathsf{Map}} \vDash T_{\mathsf{Map}} \sqsubseteq M_{\mathsf{Map}}$. $M_{\mathsf{Map}}$ is almost like $S_{\mathsf{Map}}$ module but contains several additional instructions to lubricate step-wise refinement. Executing `assume`$(P)$ in $M_{\mathsf{Map}}$ allows one to assume the fact $P$ in the lower refinement and imposes one to prove $P$ in the upper refinement, and vice versa for `assert`$(P)$. Similarly, $x?$ allows one to assume that $x$ (of an option type) has some value and extracts such a value in the lower refinement, and imposes one to prove that $x$ has some value in the upper refinement.

The key to establishing this refinement is the *condition* specified in $C_{\mathsf{Map}}$: *uninit* ensures that `init` gets called at most once. This in turn ensures that the contents of `data` and `map` always coincide, and the condition in `assert`(`size = ` None) holds true. The rest is straightforward.

Note well that, in this verification, we are enjoying the benefits of refinement (gradual verification) and unary separation logic (conditional verification) **at the same time**. In other words, conditional refinement is not, *e.g.*, disjoint union of two techniques: it really *fuses* two techniques.

## 2.2 Relational Separation Logic and Fusing All Three Styles in CCR 2.0

Now, we bring *relational* separation logic to the table: we discuss its relationship with CCR 1.0 and show how CCR 2.0 fuses the benefits of *all three* styles, yielding a novel verification strategy.

***Relational separation logic.*** *Relational* separation logics [Frumin et al. 2018, 2021; Gäher et al. 2022; Mansky et al. 2020; Sammler et al. 2019, 2023; Timany et al. 2017; Turon et al. 2013; Yang 2007], are a separation-logic-based proof technique to establish refinement.

The compositional notion of correctness here is a Hoare *quadruple* of the form: $\{P\}\, T \leq S\, \{Q\}$. The Hoare quadruple reads as: given precondition $P$, $T$ refines (simulates) $S$, and if they terminate, the resulting state satisfies $Q$. Just like Hoare triples, Hoare quadruples are *conditional* and offers *frame rule* for modular reasoning on memory locations. The key difference between Hoare triples and Hoare quadruples is manifested in their primitive connectives: in unary logic there is only one memory and one points-to connective ($\mapsto$) for that memory, whereas in relational logic there are two different memories (target and source) and two different points-to connectives ($\mapsto^{\text{tgt}}$, $\mapsto^{\text{src}}$).

The adequacy theorem of relational separation logics turns a Hoare Quadruple into a vanilla refinement judgment when the conditions $P$ and $Q$ are certain trivial conditions.

Here, we concretely consider the relational separation logic used in DimSum [Sammler et al. 2023] for following reasons. Perhaps the most well-known relational separation logic at this point would be Simuliris, but the notion of refinement in Simuliris is syntactically defined for a specific language. DimSum advances this and provides the same logic (except for concurrency) in a very close setting with CCR: *i.e.*, on top of the semantic and language-agnostic refinement.

Specifically, DimSum defines a *relational separation logic condition*, $\mathcal{J}$, satisfying (simplified):

$$\frac{\forall f \in \text{String}.\ (T[f] = b_t \wedge S[f] = b_s) \implies \{\top\}\, b_t \leq b_s\, \{\top\}}{\mathcal{J} \vDash T \sqsubseteq S} \quad (\textsc{adq-dimsum-j})$$

saying that, if every function satisfies the Hoare Quadruple with *trivial* conditions, one can get a refinement with condition $\mathcal{J}$ that can be vertically composed. $\mathcal{J}$ allows one to use relational separation logic in its internal reasoning, but its pre/postconditions are (expectedly, from the premise of the theorem) trivial.

***Complementary benefits of* CCR 1.0 *and relational separation logic.*** While both CCR 1.0 and relational separation logic spans separation logic and refinement, they are different techniques and have complementary benefits.

- Relative benefit of CCR 1.0 is that:
  In CCR 1.0, one can establish (unary) conditions in a *gradually* via vertical composition.

As we have seen in our running example, CCR 1.0 allows establishing the final condition $E_{\mathsf{Map}}$ gradually (first $C_{\mathsf{Map}}$ and then $D_{\mathsf{Map}}$). In other words, CCR 1.0 fuses both refinement and separation logic and thus enjoys their benefits **at the same time** (column **VC** × **USL**). This is not the case in relational separation logic: the benefits of these two are kept **separate**. On one hand, the Hoare quadruple judgment is conditional and satisfies the frame rule of separation logic, but is *not* vertically composable. On the other hand, when one applies the adequacy theorem to the Hoare quadruple turns it to refinement, the resulting refinement is vertically composable but is *un*-conditional.

- Relative benefit of relational separation logic is that:
  It supports arbitrary memory-changing translations (with two connectives, $\mapsto^{\text{tgt}}$ and $\mapsto^{\text{src}}$).

CCR 1.0 has difficulty in supporting arbitrary translations that change memory (global state in general), inherited from the underlying vanilla refinement and unary separation logic. All examples verified in CCR 1.0 follows the diagram below (conditions are omitted) inspired by Gu et al. [2015].

$$\overbrace{T \sqsubseteq M_0 \sqsubseteq \cdots \sqsubseteq M_m}^{\text{memory unchanging}} \underbrace{\sqsubseteq}_{\text{memory abstraction}} \overbrace{N_0 \sqsubseteq \cdots \sqsubseteq N_n \sqsubseteq S}^{\text{memory unchanging}}$$

It has a *single* "memory abstraction" translation ($M_m$ to $N_0$) that migrates the data in memory (global state) to a private state of the module: it has to replace all memory operations with corresponding operations for private states. In Fig. 2, this translation is $T_{\mathsf{Map}} \sqsubseteq M_{\mathsf{Map}}$. Note that all other translations are not changing the memory. Thus, all the modules on the left ($T$ to $M_m$) has exactly the same memory, and all the modules on the right ($N_0$ to $S$) does not have any memory operation.

This seemingly arbitrary limitation is actually crucial in CCR 1.0. Inherited from *unary* separation logic, CCR 1.0 [2] provides a unary $\mapsto$ connective. Allowing arbitrary memory changes would render the meaning of this connective ambiguous (*i.e.*, is it source or target memory?). The diagram above eliminates this ambiguity by ensuring either both source and target memory remain identical ("memory unchanging"), or memory operations occur only in the target ("memory abstraction").

While CCR 1.0's verification following the diagram above is reasonably powerful, the limitations are becoming practical problems as the framework grows.

***A motivating example.*** Consider another implementation of `Map`, $U_{\mathsf{Map}}$, that expects integer (32 bit) values and, in return, uses a more compact memory allocation, `calloc(sz * 4)` in `init`. Recall that $T_{\mathsf{Map}}$ is accepting long (64 bit) values and is using `calloc(sz * 8)`. Then, $U_{\mathsf{Map}}$ behaves the same as $T_{\mathsf{Map}}$ under the following condition $B_{\mathsf{Map}}$:

$$\forall \mathsf{k}\ \mathsf{v}.\ \{\mathsf{v} \in [-2^{32}, 2^{32})\}\ \mathtt{set(k,v)}\ \{\top\}$$

When this new implementation $U_{\mathsf{Map}}$ is added, we would like to reuse existing proof as much as possible. Generally speaking, when we have $T_{\mathsf{Map}} \sqsubseteq \ldots \sqsubseteq M_{\mathsf{Map}} \sqsubseteq \ldots \sqsubseteq S_{\mathsf{Map}}$ and a new implementation $U_{\mathsf{Map}}$ is added, the user should be able to pick whatever existing module $X$ that is closest to $U_{\mathsf{Map}}$ and provide just $U_{\mathsf{Map}} \sqsubseteq X$ as an additional proof, reusing $X \sqsubseteq S_{\mathsf{Map}}$.

In our running example, the closest module to $U_{\mathsf{Map}}$ is $T_{\mathsf{Map}}$, so we would want to prove $U_{\mathsf{Map}} \sqsubseteq T_{\mathsf{Map}}$ and compose as follows for some conditions ($\ldots$):

$$\ldots \vDash U_{\mathsf{Map}} \sqsubseteq T_{\mathsf{Map}} \quad \wedge \quad \ldots \vDash T_{\mathsf{Map}} \sqsubseteq M_{\mathsf{Map}} \quad \wedge \quad \ldots \vDash M_{\mathsf{Map}} \sqsubseteq S_{\mathsf{Map}} \quad \Longrightarrow \quad \ldots \vDash T_{\mathsf{Map}} \sqsubseteq S_{\mathsf{Map}}$$

[2]Like Iris, CCR has a base framework which is language-agnostic and does not have a notion of memory. On top of the base framework, CCR provides a default language, its notion of memory, and points-to connective. We are referring to those.

This verification *is* supported by CCR 2.0, but *not* by CCR 1.0 nor relational separation logic. First, it violates the verification pattern supported by CCR 1.0 for two reasons: (i) there are *two* translations that change memory ($U_{\text{Map}} \sqsubseteq T_{\text{Map}}$ and $T_{\text{Map}} \sqsubseteq M_{\text{Map}}$), and (ii) $U_{\text{Map}} \sqsubseteq T_{\text{Map}}$ is changing the parameters of memory operations, not "removing" memory operations. Second, as we have already discussed in §2.1, no relational separation logic supports the verification and composition of $\ldots \vDash T_{\text{Map}} \sqsubseteq M_{\text{Map}}$ and $\ldots \vDash M_{\text{Map}} \sqsubseteq S_{\text{Map}}$.

Let us see how this example is supported in CCR 2.0 in more detail.

***Fusing all three styles in* CCR 2.0.** We begin with defining $\mathcal{J}$ as in DimSum. CCR's model for conditional refinement is more sophisticated than the one in DimSum, and as a result, we have the following more powerful adequacy theorem:

$$\frac{\forall f \in \text{String}.\ (T[f] = b_t \wedge S[f] = b_s) \implies \{P_c * \mathcal{I}\ \sigma_t\ \sigma_s\}\ (\sigma_t, b_t) \le (\sigma_s, b_s)\ \{\sigma_t\ \sigma_s.\ Q_c * \mathcal{I}\ \sigma_t\ \sigma_s\}}{\mathcal{J} * C \vDash T \sqsubseteq S} \quad \text{(ADQ-CCR-J)}$$

where $\sigma_t, \sigma_s$ are module-private states and the orange coloring is just for readability. $\mathcal{I}$ is a relational invariant relating module-private states. Now, note the prevalence of $C$: which could be arbitrary conditions, like, conditions on private states of other modules. This is in contrast to ADQ-DIMSUM-J where it had to be *un*-conditional. This additional capability in ADQ-CCR-J is precisely the key that allows us to seamlessly *fuse* the benefit of CCR 1.0 and relational separation logic.

To see how $\mathcal{J}$ can be seamlessly fused into CCR 1.0's verification, let us first revisit Fig. 2.

$$\dfrac{\dfrac{\dfrac{\mathcal{J} * C_{\text{Map}} \vDash T_{\text{Map}} \sqsubseteq M_{\text{Map}} \qquad \varepsilon * D_{\text{Map}} \vDash M_{\text{Map}} \sqsubseteq S_{\text{Map}}}{(\mathcal{J} * C_{\text{Map}}) \ltimes (\varepsilon * D_{\text{Map}}) \vDash T_{\text{Map}} \sqsubseteq S_{\text{Map}}}\ (1)}{(\mathcal{J} \ltimes \varepsilon) * (C_{\text{Map}} \ltimes D_{\text{Map}}) \vDash T_{\text{Map}} \sqsubseteq S_{\text{Map}}}\ (2)}{\mathcal{J}^+ * E_{\text{Map}} \vDash T_{\text{Map}} \sqsubseteq S_{\text{Map}}}\ (3)$$

(1) We add $\mathcal{J}$ to the lower refinement (which changes memory) to leverage relational separation logic, while using $\varepsilon$ for the upper refinement (which does not change memory), leaving its verification unchanged. The $*$ between sets of conditions simply computes point-wise (for each function name) separating conjunction for both pre/postconditions. (2) Then, we apply the following key compositionality theorem developed in CCR 2.0 (and was not admissible in CCR 1.0)

$$\frac{C \vDash T \sqsubseteq M \qquad D \vDash M \sqsubseteq S}{C \ltimes D \vDash T \sqsubseteq S} \quad \text{(Enhanced VC)}$$

named Enhanced VC (enhanced vertical compositionality) to compose these two refinements. The vertical composition operator for conditions, $\ltimes$, differs from $*$ in an important respect: it is not commutative. The intuition here is, $\ltimes$ is introduced only as a consequence of applying Enhanced VC. When one has $C \ltimes D$, $C$ concerns the lower refinement (before composition) and $D$ concerns the upper refinement (before composition). $*$ is the usual separating conjunction. $\varepsilon$ is an identity element for $\ltimes$.

The most non-trivial part above is (3), where we apply an "interchange" theorem (Theorem 9). When $C_{\text{Map}}$ and $D_{\text{Map}}$ are *unary* conditions and the way relational conditions $\mathcal{J}$ and $\varepsilon$ change memory does not affect the unary conditions, we can interchange those operators. To make sense of this rewrite, note that $\mathcal{J}$ and $C_{\text{Map}}$ still relate to the lower refinement regardless of this interchange (appearing on the left of $\ltimes$), while $\varepsilon$ and $E_{\text{Map}}$ relate to the upper refinement.

Finally, we simply rewrite $C_{\text{Map}} \ltimes D_{\text{Map}}$ to $E_{\text{Map}}$ (definitionally equal) and $\mathcal{J} \ltimes \varepsilon$ to $\mathcal{J}^+$. $\mathcal{J}^+$ is an abbreviation for a finite, non-zero repetition of $\mathcal{J}$ (*i.e.*, $\mathcal{J} \ltimes \ldots \ltimes \mathcal{J}$) inspired by CompCertO [Koenig and Shao 2021]. Intuitively, those $\mathcal{J}$ in the conditional refinement judgment is innocuous, and we do not care how many times they are repeated (discussed further in §3).

Now, let us consider adding $U_{\mathsf{Map}}$. Unlike CCR 1.0, where the left judgment was *false*, the right conditional refinement is now provable—leveraging relational separation logic.

$$B_{\mathsf{Map}} \vDash U_{\mathsf{Map}} \not\sqsubseteq T_{\mathsf{Map}} \qquad\qquad \mathcal{J} * B_{\mathsf{Map}} \vDash U_{\mathsf{Map}} \sqsubseteq T_{\mathsf{Map}}$$

This refinement can be composed with existing $\mathcal{J}^+ * E_{\mathsf{Map}} \vDash T_{\mathsf{Map}} \sqsubseteq S_{\mathsf{Map}}$ exactly the same as before:

$$\cfrac{\cfrac{\cfrac{\mathcal{J} * B_{\mathsf{Map}} \vDash U_{\mathsf{Map}} \sqsubseteq T_{\mathsf{Map}} \qquad \mathcal{J}^+ * E_{\mathsf{Map}} \vDash T_{\mathsf{Map}} \sqsubseteq S_{\mathsf{Map}}}{(\mathcal{J} * B_{\mathsf{Map}}) \ltimes (\mathcal{J}^+ * E_{\mathsf{Map}}) \vDash U_{\mathsf{Map}} \sqsubseteq S_{\mathsf{Map}}}\ (1)}{(\mathcal{J} \ltimes \mathcal{J}^+) * (B_{\mathsf{Map}} \ltimes E_{\mathsf{Map}}) \vDash U_{\mathsf{Map}} \sqsubseteq S_{\mathsf{Map}}}\ (2)}{\mathcal{J}^+ * (B_{\mathsf{Map}} \ltimes E_{\mathsf{Map}}) \vDash U_{\mathsf{Map}} \sqsubseteq S_{\mathsf{Map}}}\ (3)$$

Finally, as an interesting side note, we discuss how HQ-VCOMP—where both unary predicates (like $- \mapsto_{\mathsf{Map}} -$ and $\mathit{uninit}$) and relational predicates (like $\mapsto^{\mathrm{tgt}}$ and $\mapsto^{\mathrm{src}}$) are supported—fits within this framework. Here, we follow a common pattern that we convert an intensional definition to a more extensional definition satisfying better compositional properties (from syntactic types to semantic types [Timany et al. 2024] or from compiler proof techniques to RUSC [Song et al. 2019]). Specifically, we convert Hoare Quadruple to conditional refinement judgment $\mathcal{J}^+ * C \vDash T \sqsubseteq S$ via ADQ-CCR-J. Then, this judgment enjoys vertical compositionality via ENHANCED VC.

### 2.3 Structure of the Rest of the Paper

While we have used relational separation logic as the flagship example for CCR 2.0, the key technical challenges are not in relational separation logic itself. The key challenges lie in evolving the foundation of CCR 1.0 into CCR 2.0 (column **VC** × **USL** and column **SL Interface**): these are substantial improvements on their own, and are also crucial to support relational separation logic. Henceforth, we simply say Problem **I** and **II** to refer to these issues.

During early development, we were unaware of fundamental flaws hiding in CCR 1.0's underlying model. Only after numerous failures on top of a flawed model, we were able to uncover highly non-trivial counterexamples (for **I** and **II**, each) which clearly manifested fundamental issues in the model of CCR 1.0. Guided by these counterexamples, we came up with solutions (for **I** and **II**, each) and updated the model, resulting in CCR 2.0.

Building upon the foundation of CCR 2.0, development of $\mathcal{J}$ and its Hoare Quadruple interface took less than a month without a significant challenge.

The rest of the paper is organized as follows:

- In §3, we discuss more backgrounds.
- In §4, we articulate fundamental problems **I** and **II** and give high-level overview of solutions.
- In §4.1 (Solution **I**), we define an algebra on conditions, *Generalized Conditional Refinements* which admit the *Enhanced Vertical Compositionality* (ENHANCED VC) theorem.
- In §4.2 (Solution **II**), we define sepSim and prove its key property: the *Vertical Framing*.
- In §5, we give formal definitions including the updated model in CCR 2.0 and the *Strong Update Modality* that we introduced for sepSim.
- In §6, we give an instantiation of CCR 2.0 for a tiny language and its (relational) logic with $\mathcal{J}$.
- In §7, we discuss interesting examples including the counterexamples for CCR 1.0's model.

While our results are centered around CCR, we believe its designs, algebra, and counterexamples will also benefit future researches on wrapper-based approaches and conditional refinements.

## 3 BACKGROUND

In this section, we briefly cover some essential background for the rest of the paper.

$$\begin{aligned}
\textbf{sProp} &\triangleq \Sigma \to \textbf{Prop} \\
c, d, e, fr \in \text{Cond} &\triangleq \{(\mathsf{X}, \mathsf{P}, \mathsf{Q}) \mid \mathsf{X} \in \textbf{Set} \land \mathsf{P}, \mathsf{Q} \in \mathsf{X} \to \text{Any} \to \text{Any} \to \textbf{sProp}\} \\
C, D, E, FR \in \text{Conds} &\triangleq \textbf{String} \to \text{Cond}
\end{aligned}$$

Fig. 3. Definitions of separation logic conditions and its collection.

***Refinement.*** The notion of refinement that CCR is based on is *contextual refinement.*

$$\begin{aligned}
T \sqsubseteq_{\text{closed}} S &\triangleq \text{Beh}(T) \subseteq \text{Beh}(S) \\
T \sqsubseteq \quad\quad S &\triangleq \forall Ctx.\, (Ctx \oplus T) \sqsubseteq_{\text{closed}} (Ctx \oplus S)
\end{aligned}$$

First, closed program refinement ($\sqsubseteq_{\text{closed}}$) is simply defined as set inclusion between sets of possible behaviors (*i.e.*, set of observable traces) of the given whole program [Leroy 2006]. Then, contextual refinement $\sqsubseteq$ is defined as a closed program refinement under arbitrary closing context, $Ctx$.

***Separation logic propositions.*** Generalized from the traditional separation logic, which handles ownership on a heap, modern separation logics [Appel 2014; Jung et al. 2015; Ley-Wild and Nanevski 2013] support a *user-defined* notion of ownership. Such a user-defined notion of ownership is formalized using a variant of PCM (partial commutative monoids) [Pym et al. 2004] and the whole logic is parameterized with a user-defined PCM.

As a base logic, we use a non-step-indexed version of Iris (conceptually, the same one used in Simuliris). That is, we use non-step-indexed Iris resource algebra (which is almost the same as PCM). As shown in Fig. 3, separation logic propositions **sProp** in CCR is defined as $\Sigma \to$ **Prop** where $\Sigma$ is a user-given resource algebra. Logical connectives are defined on this **sProp** space, exactly the same way as in non-step-indexed Iris.

Operators like separating conjunction ($*$) and magic wand ($-\!\!*$), quantifiers ($\forall$, $\exists$), logical operators ($\land$, $\lor$, $\implies$), entailments ($\vdash$), and their rules are completely standard and omitted in this paper.

Iris-specific operators that are relevant in this paper are as follows. $\ulcorner P \urcorner$ lifts a **Prop** into **sProp**. By convention, when applying a resource to a **sProp**, we write $[\![P]\!]a$ instead of $P(a)$. A proposition $\boxed{a}$ denotes an ownership over the underlying resource, $a$. An update modality $\dot{\Rrightarrow}$ in the goal allows one to *update* $\boxed{a}$ in the premise to $\boxed{b}$ as long as $a \rightsquigarrow b$ (definition given in §5) holds.

***Separation logic conditions in*** CCR. Cond in Fig. 3 is a separation logic condition for a single function in CCR. It comprises an auxiliary variable [Kleymann 1999], $\mathsf{X}$, and pre/post-conditions, $\mathsf{P}/\mathsf{Q}$. Auxiliary variables are those universally quantified variables that are used in both $\mathsf{P}/\mathsf{Q}$.

Note that $\mathsf{P}$ and $\mathsf{Q}$ take *two* Any-type parameters (Any can be seen as a union of all values we use[3]). These two parameters (intuitively) represent source and target values respectively, and are used in defining Hoare quadruples. For unary conditions, we implicitly impose the source and target values to be identical. For example, the Hoare triple notation from Fig. 2 unfolds as follows:

$$\forall x \in X.\, \{P\,x\,a\}\; b(a)\; \{r.\, Q\,x\,r\} \quad \textit{means} \quad \{X,\;\; \lambda\, a_s\, a_t.\, \ulcorner a_s = a_t \urcorner * P\,x\,a_s,\;\; \lambda\, r_s\, r_t.\, \ulcorner r_s = r_t \urcorner * Q\,x\,r_s\}$$

Conds is simply a collection of Cond, indexed with function name.

***Conditional Refinements in*** CCR 1.0. One of the key ideas behind CCR is to define the notion of conditional refinement on top of the vanilla, *un*-conditional refinement:

$$C \vDash T \sqsubseteq S \quad \triangleq \quad T \sqsubseteq \lfloor S \rfloor_C$$

Here, the **wrapper** on the source side ($\lfloor - \rfloor_S$) adjusts the *operational semantics* of the given module so that the conditions on $C$ are *dynamically* enforced—exactly how CCR 1.0 achieves it is where the main complexity is, and we keep it abstract here.

[3] In Rocq, it is simply implemented as a dependent pair type, $\Sigma_{t:\mathsf{Type}_u} t$, for a certain universe $u$.

This definition allows one to piggyback on the vertical compositionality of the underlying vanilla refinement. Indeed, CCR 1.0 did not have Enhanced VC theorem, and instead relied on this low-level notion of vertical composition. We will soon discuss why this was problematic.

***Wrapper Elimination Theorem.*** The adequacy theorem of CCR is stated as a *Wrapper Elimination Theorem (WET).*

$$\frac{\forall i \in \{1, 2, \ldots, n\}.\ C \vDash T_i \sqsubseteq S_i \qquad T_1 \oplus \ldots \oplus T_n \text{ is a closed program.}}{T_1 \oplus \ldots \oplus T_n \sqsubseteq_{\text{closed}} S_1 \oplus \ldots \oplus S_n} \quad \text{(WET)}$$

This theorem states that, given a closed program, when every module $T_i$ establishes refinement against their specifications $S_i$ under the same conditions, $C$, we can "cancel out" those conditions and obtain whole-program vanilla refinement as the final result. For clarity: having "the same" conditions $C$ for every module-wise refinement means that the caller and callee agree upon the specifications, not that every function has an identical specification (recall that $C$ is a collection of function-wise pre/postconditions).

Now we briefly discuss why $\mathcal{J}^+$ is "innocuous" and why the exact number of repetitions of $\mathcal{J}$ do not matter. Following CompCertM [Song et al. 2019], we leverage the "self-relatedness" property of modules. Any module written in the default programming language provided by CCR are trivially self-related by $\mathcal{J}$: *i.e.*, $\mathcal{J} * \varepsilon \vDash T \sqsubseteq T$ holds. Consider a scenario where we have $\mathcal{J}^3 * C \vDash T_1 \sqsubseteq S_1$ and $\mathcal{J}^4 * C \vDash T_2 \sqsubseteq S_2$. This appears problematic since $\mathcal{J}^3$ and $\mathcal{J}^4$ differ, preventing the application of WET. However, we can equalize the number of $\mathcal{J}$ occurrences by exploiting self-relatedness:

$$\cfrac{\cfrac{\cfrac{\mathcal{J} * \varepsilon \vDash T_1 \sqsubseteq T_1 \qquad \mathcal{J}^3 * C \vDash T_1 \sqsubseteq S_1}{(\mathcal{J} * \varepsilon) \ltimes (\mathcal{J}^3 * C) \vDash T_1 \sqsubseteq S_1}}{(\mathcal{J} \ltimes \mathcal{J}^3) * (\varepsilon \ltimes C) \vDash T_1 \sqsubseteq S_1}}{\mathcal{J}^4 * C \vDash T_1 \sqsubseteq S_1}$$

## 4 PROBLEMS OF CCR 1.0 AND SOLUTIONS IN CCR 2.0

We now examine problems **I** and **II** in CCR 1.0 and their solutions in CCR 2.0.

***Problem I: Limited proof reuse and modularity.*** Recall that in Fig. 2, we established $E_{\text{Map}} \vDash T_{\text{Map}} \sqsubseteq S_{\text{Map}}$ by composing the following two refinements:

$$C_{\text{Map}} \vDash T_{\text{Map}} \sqsubseteq M_{\text{Map}} \qquad D_{\text{Map}} \vDash M_{\text{Map}} \sqsubseteq S_{\text{Map}}$$

This seemingly simple composition does work in CCR 2.0, but it does *not* work in CCR 1.0!

The reason is as follows. If we unfold the definition of conditional refinement in CCR 1.0:

$$T_{\text{Map}} \sqsubseteq \lfloor M_{\text{Map}} \rfloor_{C_{\text{Map}}} \qquad M_{\text{Map}} \sqsubseteq \lfloor S_{\text{Map}} \rfloor_{D_{\text{Map}}}$$

we have these two refinements, and the module in the middle does not coincide with each other ($\lfloor M_{\text{Map}} \rfloor_{C_{\text{Map}}}$ in the lower refinement and $M_{\text{Map}}$ in the upper refinement). CCR 1.0 relied on transitivity of underlying vanilla refinement to vertically compose conditional refinements, and transitivity does not apply here.

In order to use the transitivity of underlying refinement, CCR 1.0 had to set up the following two refinements instead from the beginning:

$$C_{\text{Map}} \vDash T_{\text{Map}} \sqsubseteq M_{\text{Map}} \qquad E_{\text{Map}} \vDash \lfloor M_{\text{Map}} \rfloor_{C_{\text{Map}}} \sqsubseteq S_{\text{Map}}$$

where the upper refinement unnecessarily gets exposed to the conditions that are just relevant for the lower refinement ($C_{\text{Map}}$ on the left side, and use of $E_{\text{Map}}$ instead of $D_{\text{Map}}$ on the right side).

While less ideal, the above was not a show-stopper for the verifications in CCR 1.0. However, when we consider adding a new implementation, $U_{\mathsf{Map}}$, the problem becomes severe. Recall that we proved the following two refinements and composed:

$$B \vDash U_{\mathsf{Map}} \sqsubseteq T_{\mathsf{Map}} \qquad\qquad E \vDash T_{\mathsf{Map}} \sqsubseteq S_{\mathsf{Map}}$$

where we abbreviate $\mathcal{J} * B_{\mathsf{Map}}$ to $B$ and $\mathcal{J}^+ * E_{\mathsf{Map}}$ to $E$ (the exact condition is not important here). For the same reason as above, these two refinements do not compose. However, unlike above, we cannot just avoid this issue by adjusting the upper refinement to $F \vDash \lfloor T_{\mathsf{Map}} \rfloor_B \sqsubseteq S_{\mathsf{Map}}$ for some $F$. We need *both* $E \vDash T_{\mathsf{Map}} \sqsubseteq S_{\mathsf{Map}}$ (for implementation $T_{\mathsf{Map}}$) and $F \vDash \lfloor T_{\mathsf{Map}} \rfloor_B \sqsubseteq S_{\mathsf{Map}}$ (for implementation $U_{\mathsf{Map}}$), and CCR 1.0 does not provide any way to accommodate proof reuse between the two.

Generally speaking, when we have $T_{\mathsf{Map}} \sqsubseteq \ldots \sqsubseteq M_{\mathsf{Map}} \sqsubseteq \ldots \sqsubseteq S_{\mathsf{Map}}$ and a new implementation $U_{\mathsf{Map}}$ is added, the user should be able to pick whatever existing module $X$ that is closest to $U_{\mathsf{Map}}$ and provide just $U_{\mathsf{Map}} \sqsubseteq X$ as an additional proof, reusing $X \sqsubseteq S_{\mathsf{Map}}$. While we considered just one case here (where $X$ is $T_{\mathsf{Map}}$), CCR 1.0 suffers from the same limitation regardless of what $X$ we choose.

To resolve this, we need a more separable notion of conditional refinement and a better, higher-level vertical compositionality.

***Solution I: Enhanced vertical compositionality.*** As hinted before, in CCR 2.0, we solve this issue with the following higher-level vertical compositionality theorem for conditional refinements called *enhanced vertical compositionality*:

$$\frac{C \vDash T \sqsubseteq M \qquad D \vDash M \sqsubseteq S}{C \ltimes D \vDash T \sqsubseteq S} \qquad (\textsc{Enhanced VC})$$

This theorem is "enhanced" from the transitivity of underlying vanilla refinements ($\sqsubseteq$): putting $\varepsilon$ for both $C$ and $D$ yields the underlying transitivity.

As an analogy, this resembles thread-wise reasoning in concurrent separation logic. There, the ownership is split and distributed to each *thread*, and each thread can be reasoned modularly. Here, the ownership ($C \ltimes D$) is split and distributed to each *refinement*, and each refinement can be reasoned modularly. We dubbed this form of modular reasoning a **Vertical Framing**, contrasting it with the usual framing in separation logic (which we call a *Horizontal Framing* in this regard).

The key challenge in Solution **I** is that there is a subtle counterexample that prevents CCR 1.0's definition of conditional refinement from satisfying Enhanced VC. We fix this issue by slightly adjusting the definition, dubbed *generalized conditional refinements*, while maintaining full usability.

***Problem II: Model-level reasoning for separation logic conditions.*** Now we are ready to look at the problem **II**. While CCR 1.0 defined the wrapper, it involves sophisticated low-level details, requiring explicit manipulation of resources (the model of separation logic; we henceforth use jade for model-level objects). They did provide automation for resource manipulations, but (i) it worked only when the source side is wrapped (*i.e.*, $T \sqsubseteq \lfloor S \rfloor_C$), and (ii) still exposed the resources to the user.

When the target side is also wrapped, the proof is inherently much more complex. To get a glimpse of what such proofs look like in CCR 1.0, we took an excerpt from Song et al. [2023]. To establish $\lfloor M_{\mathsf{Map}} \rfloor_{C_{\mathsf{Map}}} \sqsubseteq \lfloor S_{\mathsf{Map}} \rfloor_{E_{\mathsf{Map}}}$, they proved $\lfloor M_{\mathsf{Map}} \rfloor_{C_{\mathsf{Map}}} \precsim \lfloor S_{\mathsf{Map}} \rfloor_{E_{\mathsf{Map}}}$ directly using a low-level simulation technique, diving into the wrapper's definition. Below is the sketch for the `init` case:

> *...(omitted)... At the end of the function, we execute* `*assert` *in the target, which gives us the updated $\mu'_M$ and the validity condition $\checkmark(\mu'_M \cdot \mathtt{ctx}_M)$ (we ignore the return resource since the target post is $\top$). We then execute* `*assert` *in the source by picking $\rho_S$ to be the resource satisfying $\ast_{k\in[0,\mathtt{size})}\ \mathtt{k} \mapsto_{\mathsf{Map}} \mathtt{0}$, updating $\mu'_S$ to be $\mu'_M \cdot \mathit{uninit}_1 \cdot \bullet(\mathtt{map}_S\langle 0 : \mathtt{size}\rangle)$, and proving the validity condition $\checkmark(\mu'_S \cdot \rho_S \cdot \mathtt{ctx}_S)$. This validity condition is implied by the validity condition from the target and the fact that the allocation of $\ast_{k\in[0,\mathtt{size})}\ \mathtt{k} \mapsto_{\mathsf{Map}} \mathtt{0}$ (and corresponding update to $\bullet(\mathtt{map}_S\langle 0 : \mathtt{size}\rangle)$) are frame-preserving updates. Finally, we have to reestablish the invariant, which is straightforward since $\mathtt{map}_M$ and $\mathtt{map}_S$ are not modified and $\mu'_S$ contains $\mu'_M$, $\bullet(\mathtt{map}_S\langle 0 : \mathtt{size}\rangle)$, and $\mathit{uninit}_1$.*

We will not delve into the details of this proof: The point here is that this proof exposes model-level details of the wrapper (the resources $\mu'_M, \mathtt{ctx}_M, \rho_M$, and $\mu'_S, \mathtt{ctx}_S, \rho_S$) to the user, where all these do *not* even appear in the original modules ($M_{\mathsf{Map}}$ and $S_{\mathsf{Map}}$) and conditions ($C_{\mathsf{Map}}$ and $D_{\mathsf{Map}}$); they are *internal details* of the wrapper that should be hidden from the user.

This proof strategy is not only complex but also ad hoc, specific to this example. CCR 1.0 lacks a *general* high-level reasoning principle, which required (as it turns out) changes in the model.

As an analogy, consider Iris: while even a simple magic wand operator has a sophisticated model, Iris base logic does the heavy lifting by hiding model-level details and providing abstract logical rules. This enables users to develop program logic for their chosen language on top of the language-agnostic Iris base logic. Similarly, CCR 2.0 hides the detailed model of the wrapper and provides logical rules to users.

Coming back to relational separation logic: Hoare Quadruples in Simuliris or DimSum are also defined using logical connectives defined in Iris, hiding all the model-level details. For $\mathcal{J}$ to provide the same user-interface, expressed as logical rules, it is crucial for the underlying framework to provide logical rules for the primitives it offers.

***Solution II: A high-level reasoning principle.*** To address this problem, we introduce a new proof technique, a separation-logic-simulation (shortly, sepSim) $C \mid D \Vdash T \precsim^{*} S$, (roughly) satisfying:

$$C \mid D \Vdash T \precsim^{*} S \implies \lfloor T \rfloor_C \sqsubseteq \lfloor S \rfloor_D \qquad \text{(ADEQUACY-SEPSIM)}$$

First, note that separation logic conditions ($C$ and $D$) and the program code ($T$ and $S$) are clearly separated. $C \mid D \Vdash T \precsim^{*} S$ behaves the same as a usual simulation technique[4], $T \precsim S$, and then gives logic-level proof obligation with respect to $C$ and $D$ at each call site. In other words, in sepSim, the goal is a usual simulation technique that only concerns *user-written* codes ($T \precsim S$), completely hiding model-level details (those $\mu, \mathtt{ctx}, \rho$) and exposes only the separation-logic proof obligation that these instructions are meant to encode to the user.

To get a better sense of what sepSim looks like, consider the following simple case:

$$\frac{\forall v.\, P_s \twoheadrightarrow (P_t * f \mapsto (P_t, Q_t) \mid f \mapsto (P_s, Q_s) \Vdash b_t(v) \precsim^{*} b_s(v)\ \{Q_t \twoheadrightarrow Q_s\})}{f \mapsto (P_t, Q_t) \mid f \mapsto (P_s, Q_s) \Vdash \mathbf{def}\ f(v)\ \{b_t(v)\} \precsim^{*} \mathbf{def}\ f(v)\ \{b_s(v)\}}$$

where the function $f$ has pre/postconditions $P_t, Q_t$ in the target (and similarly for the source). Then, we get a proof obligation that gives us $P_s$ in the beginning, mandates us to provide $P_t$ out of it, and proceed with the proof with the remaining spatial propositions. After executing the body of both sides $b_t(v)$ and $b_s(v)$, one is obligated to prove the postcondition ($\{Q_t \twoheadrightarrow Q_s\}$) which gives $Q_t$ to the user, and obligates to prove $Q_s$ together with remaining spatial propositions.

Finally, unlike before, sepSim itself is a separation logic proposition and enjoys standard structural rules in Iris-style weakest preconditions including the frame rule and the bind rule. We fully utilize the Iris Proof Mode [Krebbers et al. 2017] to further streamline the proof.

[4]Specifically, sepSim is based on an advanced simulation technique recently proposed in [Cho et al. 2023].

$$
\begin{aligned}
c \equiv d &\triangleq \exists \rightharpoonup\, \in \mathsf{X}_c \to \mathsf{X}_d.\, \exists \leftharpoondown\, \in \mathsf{X}_d \to \mathsf{X}_c.\ \rightharpoonup \circ \leftharpoondown\ = \mathbf{Id}_{\mathsf{X}_d} \wedge\ \leftharpoondown \circ \rightharpoonup\ = \mathbf{Id}_{\mathsf{X}_c} \wedge \\
&\qquad (\forall x\, a_s\, a_t.\, \mathsf{P}_c\, x\, a_s\, a_t \dashv\vdash \mathsf{P}_d\, (\rightharpoonup x)\, a_s\, a_t) \wedge (\forall x\, r_s\, r_t.\, \mathsf{Q}_c\, x\, r_s\, r_t \dashv\vdash \mathsf{Q}_d\, (\rightharpoonup x)\, r_s\, r_t) \\
c * d &\triangleq \{\mathsf{X}_c \times \mathsf{X}_d, (\lambda\, (x_c, x_d)\, a_s\, a_t.\quad \mathsf{P}_c\, x_c\, a_s\, a_t * \mathsf{P}_d\, x_d\, a_s\, a_t), \\
&\qquad (\lambda\, (x_c, x_d)\, r_s\, r_t.\quad \mathsf{Q}_c\, x_c\, r_s\, r_t * \mathsf{Q}_d\, x_d\, r_s\, r_t)\} \\
\mathrm{Cond} \ni \top &\triangleq \{\mathbf{1}, (\lambda\, _\, _.\, \top), (\lambda\, _\, _.\, \top)\} \\
c \ltimes d &\triangleq \{\mathsf{X}_c \times \mathsf{X}_d, (\lambda\, (x_c, x_d)\, a_s\, a_t.\, \exists a_m.\, \mathsf{P}_c\, x_c\, a_m\, a_t * \mathsf{P}_d\, x_d\, a_s\, a_m), \\
&\qquad (\lambda\, (x_c, x_d)\, r_s\, r_t.\ \exists r_m.\, \mathsf{Q}_c\, x_c\, r_m\, r_t * \mathsf{Q}_d\, x_d\, r_s\, r_m)\} \\
\mathrm{Cond} \ni \varepsilon &\triangleq \{\mathbf{1}, (\lambda\, _\, a_s\, a_t.\, \ulcorner a_s = a_t \urcorner), (\lambda\, _\, r_s\, r_t.\, \ulcorner r_s = r_t \urcorner)\}
\end{aligned}
$$

| | |
|---|---|
| $\forall c.\quad \top * c \equiv c \equiv c * \top$ | $\forall c.\quad \varepsilon \ltimes c \equiv c \equiv c \ltimes \varepsilon$ |
| $\forall c\, d.\quad c * d \equiv d * c$ | |
| $\forall c\, d\, e.\quad c * d * e \equiv c * (d * e)$ | $\forall c\, d\, e.\quad c \ltimes d \ltimes e \equiv c \ltimes (d \ltimes e)$ |
| $\forall c\, d\, c'\, d'.\ (c \equiv c' \wedge d \equiv d') \implies c * d \equiv c' * d'$ | $\forall c\, d\, c'\, d'.\ (c \equiv c' \wedge d \equiv d') \implies c \ltimes d \equiv c' \ltimes d'$ |

| UNIT-WRAP | EQV-PROPER |
|---|---|
| $M \sqsupseteq\sqsubseteq \lfloor M \rfloor_\varepsilon$ | $C \equiv C' \Rightarrow \lfloor M \rfloor_C \sqsubseteq \lfloor M \rfloor_{C'}$ |

Fig. 4. Operators on Cond and Conds (above), and their properties (below).

sepSim simplifies the above proof excerpt a lot: all the details (shaded sentences) are hidden from the user: sepSim does the heavy lifting of model-level reasoning for the user.

The key challenge in Solution **II** is that there is a highly non-trivial counterexample that prevents CCR 1.0's model from satisfying all the desiderata of sepSim. We fix this issue by slightly adjusting the model of the wrapper while maintaining full usability.

### 4.1 Solution I: Enhanced Vertical Compositionality

In this section, we see the key idea behind solution Solution **I**.

***An algebra for Cond (and Conds).*** To begin with, in Fig. 4, we define operators on conditions (Cond) and their collections (Conds) and prove basic properties on them. First, we define an equivalence relation ($\equiv$) for Cond: $c$ and $d$ is equivalent if there is a bijection between the set of auxiliary variables ($\rightharpoonup$ and $\leftharpoondown$), and the conditions are equivalent up-to bijection.

We have two operators for Cond: $*$ and $\ltimes$. Recall that $c * d$ merges Cond at the same refinement level, while $c \ltimes d$ connects conditions between lower (c) and upper (d) refinements. Specifically, $c * d$ is a pairwise separating conjunction of conditions from $c$ and $d$, with each condition retaining its own auxiliary variables. On the other hand, $\ltimes$ is like $*$ but existentially quantifies intermediate values, $a_m$, and $c$ relates $a_t$ and $a_m$ while $d$ relates $a_m$ and $a_s$.

The $\ltimes$ operator was not present in CCR 1.0; it is newly introduced in CCR 2.0 for Enhanced VC. An astute reader might ask if we can just use $*$ in Enhanced VC. In Appendix C, with a concrete example, we show how $*$ and $\ltimes$ are different and why $*$ does not work for Enhanced VC.

Properties for these operators are shown in the middle of Fig. 4. Both operators have identity elements: $\top$ for $*$ and $\varepsilon$ for $\ltimes$. A unit $\top$ has a unit set (**1**) as the type of the auxiliary variable and has trivial conditions. $\varepsilon$ is like $\top$ but equates source and target values. $*$ is commutative and associative, whereas $\ltimes$ is only associative. Both operators respect the equivalence relation $\equiv$.

This algebra on Cond can be lifted to Conds (pointwise), and we overload notations accordingly.

***Enhanced vertical compositionality.*** Now, we are ready to revisit the rule Enhanced VC. As a starting point, we connect the condition algebra with refinements with unit-wrap and eqv-proper, shown in Fig. 4. Specifically, using unit-wrap, we can always wrap unwrapped modules with $\varepsilon$. This allows us to work uniformly on a setting where both the source and target are wrapped—as in adequacy-sepsim—we can always wrap a module with $\lfloor - \rfloor_\varepsilon$.

With the definition of conditional refinement in CCR 1.0 ($C \vDash T \sqsubseteq S \triangleq T \sqsubseteq \lfloor S \rfloor_C$), the most straightforward way to justify the Enhanced VC rule would be by proving the following:

$$\frac{\lfloor T \rfloor_C \sqsubseteq \lfloor S \rfloor_D}{\lfloor T \rfloor_{FR \ltimes C} \sqsubseteq \lfloor S \rfloor_{FR \ltimes D}} \qquad \text{(VFRAME?)}$$

Morally, the rule should hold because $FR$ on both sides cancel each other out (precondition to precondition, and postcondition to postcondition). To get more intuition, consider the following:

$$T = (\mathbf{def}\ f_t(x)\ \{\mathbf{skip}\}) \qquad\qquad S = (\mathbf{def}\ f_s(x)\ \{\mathbf{skip}\})$$
$$C(f) = (\mathbf{1}, \lambda_\,_.\,P_t, \lambda_\,_.\,Q_t) \qquad FR(f) = (\mathbf{1}, \lambda_\,_.\,X, \lambda_\,_.\,Y) \qquad D(f) = (\mathbf{1}, \lambda_\,_.\,P_s, \lambda_\,_.\,Q_s)$$

With these simple instances, these two refinements have the following *proof obligation* written as separation logic formulae:[5]

$$P_s \twoheadrightarrow P_t * (Q_t \twoheadrightarrow Q_s) \qquad\qquad (P_s * X) \twoheadrightarrow (P_t * X) * ((Q_t * Y) \twoheadrightarrow (Q_s * Y))$$

Finally, it is easy to check that the former implies the latter.

Now, assuming the vframe? rule, we can straightforwardly prove Enhanced VC as follows:

$$\dfrac{\dfrac{\dfrac{\dfrac{\dfrac{C \vDash T \sqsubseteq M \qquad D \vDash M \sqsubseteq S}{T \sqsubseteq \lfloor M \rfloor_C \qquad M \sqsubseteq \lfloor S \rfloor_D}\ \text{By unfolding.}}{T \sqsubseteq \lfloor M \rfloor_C \qquad \lfloor M \rfloor_\varepsilon \sqsubseteq \lfloor S \rfloor_D}\ \text{By UNIT-WRAP.}}{T \sqsubseteq \lfloor M \rfloor_C \qquad \lfloor M \rfloor_C \sqsubseteq \lfloor S \rfloor_{C \ltimes D}}\ \text{By VFRAME?.}}{T \sqsubseteq \lfloor S \rfloor_{C \ltimes D}}\ \text{By transitivity of } \sqsubseteq.}{C \ltimes D \vDash T \sqsubseteq S}\ \text{By folding.}$$

However, there is one caveat in doing this—actually, the vframe? rule does not hold! A counterexample for this is subtle and technical and we postpone its discussion to §7. At the high level, in the counterexample, the target program in the first refinement $\lfloor T \rfloor_D$ is ill-formed, which unwantedly makes the first refinement trivial; the refinement is provable not with intended proof techniques but only by exploiting ill-formedness at the model-level.

To rule out such a semantically ill-formed program and fix the problem, our key idea is—taking inspiration from Birkedal et al. [2008]—to “bake-in” the vertical framing into the definition of conditional refinement. That is, we *generalize* the notion of conditional refinement as follows:

**Definition 1.** Generalized conditional refinement $C \vDash T \sqsubseteq S$ is defined as follows:

$$\forall FR.\ \lfloor T \rfloor_{FR} \sqsubseteq \lfloor S \rfloor_{FR \ltimes C}$$

First, note that it is strictly stronger than the original conditional refinement: putting $\varepsilon$ as $FR$ yields the original conditional refinement. Second, this definition now justifies Enhanced VC. The proof is almost the same as the above, except that we can now use the “baked-in” version of vframe?. Now, the only thing that is left to check is how we can establish this strengthened refinement. Have we accidentally made it overly strong that it became too hard to establish?

The answer is—fortunately—no. In the next section, we prove that this change does not affect anything for users as long as they use the framework in the intended way; *i.e.*, using sepSim.

### 4.2 Solution II: Separation Logic Simulation

In this section, we present sepSim, the main technical contribution of this paper. sepSim abstracts a tedious model-level proof of CCR 1.0 into a logic-level proof, and also supports conditions in the target properly.

[5] As we will see in the next section, there should be update modalities as well. However, the argument still holds the same.

To begin with, our notion of a module comprises three components: (i) $\texttt{funs}$, a partial function from function name to its semantics, (ii) $\texttt{init_st}$, an initial state of the module, and (iii) $\texttt{init_res}$, an initial resource of the module. Then, the adequacy theorem of sepSim establishes $\lfloor T \rfloor_C \sqsubseteq \lfloor S \rfloor_D$:

**Theorem 2** (Adequacy of sepSim). *For a given pair of modules T and S, and their wrappers, C and D, if there exists an invariant $\mathcal{I}$ and the following three conditions are met:*
(*i*) $\mathrm{dom}(\texttt{funs}_T) = \mathrm{dom}(\texttt{funs}_S)$ (*ii*) $[\![(\mathcal{I}\ \texttt{init_st}_T\ \texttt{init_st}_S) * \boxed{\texttt{init_res}_T}]\!]\texttt{init_res}_S$
(*iii*) $\forall f.\, \mathsf{sepSimFun}\ C\ D\ \mathcal{I}\ f\ \texttt{funs}_T(f)\ \texttt{funs}_S(f)$
*we have* $\lfloor T \rfloor_C \sqsubseteq \lfloor S \rfloor_D$.

In the adequacy theorem, the first condition $(i)$ says that both modules define the same function names. The second condition $(ii)$ says that the initial states satisfy the invariant $\mathcal{I}$ with the initial resource of the module, $\texttt{init_res}_S$.[6] Finally, the third condition $(iii)$ is the main proof obligation: each function should be simulated using sepSim.

Before discussing sepSim in detail, we first present its unique theorem, dubbed **Vertical Framing**.

**Theorem 3** (Vertical Framing). *The following property holds:*

$$\mathsf{sepSimFun}\ C\ D\ \mathcal{I}\ f\ b_t\ b_s \dashv\vdash (\forall FR.\, \mathsf{sepSimFun}\ (FR \ltimes C)\ (FR \ltimes D)\ \mathcal{I}\ f\ b_t\ b_s)$$

Since conditions cancel out in a vertical direction (between the source and the target), we call this vertical framing. Vertical framing is closely related to generalized conditional refinement:

Corollary 4. *Theorem 2 gives the same proof obligation for $\lfloor T \rfloor_C \sqsubseteq \lfloor S \rfloor_D$ and $\forall FR.\, \lfloor T \rfloor_{FR \ltimes C} \sqsubseteq \lfloor S \rfloor_{FR \ltimes D}$. In a special case where $C = \varepsilon$, the proof obligation for $T \sqsubseteq \lfloor S \rfloor_D$ equals that of $D \vDash T \sqsubseteq S$.*

As claimed in §4.1, *generalized* conditional refinement requires no extra proof obligation.

***Definitions and Rules of* sepSim.** We now present the formal definitions and rules, beginning with sepSimFun defined as follows:

$$\begin{aligned}
&\mathsf{sepSimFun}\ C\ D\ \mathcal{I}\ f\ b_t\ b_s \triangleq \mathbf{let}\ (c, d) := (C(f), D(f))\ \mathbf{in}\ \forall \sigma_t\ \sigma_s\ a.\\
&\quad \forall a_s\ x_d.\, (\mathcal{I}\ \sigma_t\ \sigma_s) \twoheadrightarrow (\mathsf{P}_d\ x_d\ a_s\ a) \twoheadrightarrow\\
&\quad \exists a_t\ x_c.\, (\mathsf{P}_c\ x_c\ a_t\ a) *\\
&\quad C \mid D \Vdash (\sigma_t, b_t(a_t)) \precsim (\sigma_s, b_s(a_s))\ \{\lambda \sigma_t'\ \sigma_s'\ r_t\ r_s.\, \forall r.\, (\mathsf{Q}_c\ x_c\ r_t\ r) \twoheadrightarrow (\mathcal{I}\ \sigma_t'\ \sigma_s') * (\mathsf{Q}_d\ x_d\ r_s\ r)\}
\end{aligned}$$

To understand the definition, (i) first observe that source and target conditions appear in dual positions: $(\mathsf{P}_d\ x_d\ a_s\ a)$ in negative and $(\mathsf{P}_c\ x_c\ a_t\ a)$ in positive position (the opposite for postconditions). This duality explains why Theorem 3 holds: matching conditions on source and target cancel out.

Next, (ii) consider a simple case where $\mathcal{I}$ is $\top$ and $c$ is $\varepsilon$:

$$\forall a_t\ a_s\ x_d.\, (\mathsf{P}_d\ x_d\ a_s\ a_t) \twoheadrightarrow C \mid D \Vdash (\sigma_t, b_t(a_t)) \precsim (\sigma_s, b_s(a_s))\ \{\lambda \sigma_t'\ \sigma_s'\ r_t\ r_s.\, (\mathsf{Q}_d\ x_d\ r_s\ r_t)\}$$

which reads as a usual Hoare Quadruple in separation logic: given an arbitrary auxiliary variable $x_s$ and precondition $(\mathsf{P}_d\ x_d\ a_s\ a_t)$, one should simulate the program code on both sides be able to and return with postcondition, $(\mathsf{Q}_d\ x_d\ r_s\ r_t)$.

Finally, (iii) $\mathcal{I}$ provides the usual rely/guarantee-style reasoning for the module-private states: it allows one to rely on it at the beginning of the function $(\mathcal{I}\ \sigma_t\ \sigma_s)$ and instead requires one to prove it at the end of the function $(\mathcal{I}\ \sigma_t'\ \sigma_s')$.

Now we see the rules for sepSim, shown in Fig. 5. While the complete form of sepSim is $C \mid D \Vdash (\sigma_t, b_t) \precsim (\sigma_s, b_s)\ \{Q\}$, for brevity, we omit $C \mid D \Vdash$, states $\sigma_t$ and $\sigma_s$, and postcondition $Q$ when they remain unchanged between premise and goal.

The figure presents three categories of rules.

[6] For readers familiar with Iris, $\mathcal{I}$ could be seen as a relational version of state interpretation for the private states in $T$ and $S$.

**Rules for executing physical program.**

ANGELIC-SRC
$$\frac{\forall x \in X.\, b_t \mathrel{\overset{*}{\precsim}} b_s(x)}{b_t \mathrel{\overset{*}{\precsim}} \dot{\forall} x \in X.\, b_s(x)}$$

DEMONIC-SRC
$$\frac{\exists x \in X.\, b_t \mathrel{\overset{*}{\precsim}} b_s(x)}{b_t \mathrel{\overset{*}{\precsim}} \dot{\exists} x \in X.\, b_s(x)}$$

ANGELIC-TGT
$$\frac{\exists x \in X.\, b_t \mathrel{\overset{*}{\precsim}} b_s(x)}{\dot{\forall} x \in X.\, b_t(x) \mathrel{\overset{*}{\precsim}} b_s}$$

DEMONIC-TGT
$$\frac{\forall x \in X.\, b_t \mathrel{\overset{*}{\precsim}} b_s(x)}{\dot{\exists} x \in X.\, b_t(x) \mathrel{\overset{*}{\precsim}} b_s}$$

PUT-SRC
$$\frac{(\sigma_t, b_t) \mathrel{\overset{*}{\precsim}} (\sigma_s', b_s)}{(\sigma_t, b_t) \mathrel{\overset{*}{\precsim}} (\sigma_s, \mathbf{put}(\sigma_s'); b_s)}$$

GET-SRC
$$\frac{(\sigma_t, b_t) \mathrel{\overset{*}{\precsim}} (\sigma_s, b_s(\sigma_s))}{(\sigma_t, b_t) \mathrel{\overset{*}{\precsim}} (\sigma_s, \mathbf{get} \gg\!= b_s)}$$

PUT-TGT
$$\frac{(\sigma_t', b_t) \mathrel{\overset{*}{\precsim}} (\sigma_s, b_s)}{(\sigma_t, \mathbf{put}(\sigma_t'); b_t) \mathrel{\overset{*}{\precsim}} (\sigma_s, b_s)}$$

GET-TGT
$$\frac{(\sigma_t, b_t(\sigma_t)) \mathrel{\overset{*}{\precsim}} (\sigma_s, b_s)}{(\sigma_t, \mathbf{get} \gg\!= b_t) \mathrel{\overset{*}{\precsim}} (\sigma_s, b_s)}$$

ASSUME-SRC
$$\frac{\ulcorner P \urcorner \implies b_t \mathrel{\overset{*}{\precsim}} b_s}{b_t \mathrel{\overset{*}{\precsim}} \mathtt{assume}(P); b_s}$$

ASSERT-SRC
$$\frac{\ulcorner P \urcorner \wedge b_t \mathrel{\overset{*}{\precsim}} b_s}{b_t \mathrel{\overset{*}{\precsim}} \mathtt{assert}(P); b_s}$$

ASSUME-TGT
$$\frac{\ulcorner P \urcorner \wedge b_t \mathrel{\overset{*}{\precsim}} b_s}{\mathtt{assume}(P); b_t \mathrel{\overset{*}{\precsim}} b_s}$$

ASSERT-TGT
$$\frac{\ulcorner P \urcorner \implies b_t \mathrel{\overset{*}{\precsim}} b_s}{\mathtt{assert}(P); b_t \mathrel{\overset{*}{\precsim}} b_s}$$

**Rules for executing wrapper.**

*ASSUME-SRC
$$\frac{P \mathrel{-\!\!*} b_t \mathrel{\overset{*}{\precsim}} b_s}{b_t \mathrel{\overset{*}{\precsim}} *\mathtt{assume}(P); b_s}$$

*ASSERT-SRC
$$\frac{P * b_t \mathrel{\overset{*}{\precsim}} b_s}{b_t \mathrel{\overset{*}{\precsim}} *\mathtt{assert}(P); b_s}$$

*ASSUME-TGT
$$\frac{P * b_t \mathrel{\overset{*}{\precsim}} b_s}{*\mathtt{assume}(P); b_t \mathrel{\overset{*}{\precsim}} b_s}$$

*ASSERT-TGT
$$\frac{P \mathrel{-\!\!*} b_t \mathrel{\overset{*}{\precsim}} b_s}{*\mathtt{assert}(P); b_t \mathrel{\overset{*}{\precsim}} b_s}$$

SIM-CALL
$$\frac{\begin{array}{c} c = C(f) \qquad d = D(f) \\ \forall x_c\, a.\, (\mathsf{P}_c\ x_c\ a_t\ a) \mathrel{-\!\!*} \exists x_d.\, (\mathsf{P}_d\ x_d\ a_s\ a) * (\mathcal{I}\ \sigma_t\ \sigma_s)\ * \\ (\forall \sigma_t'\ \sigma_s'\ r_s\ r.\, (\mathcal{I}\ \sigma_t'\ \sigma_s') \mathrel{-\!\!*} (\mathsf{Q}_d\ x_s\ r_s\ r) \mathrel{-\!\!*} \exists r_t.\, (\mathsf{Q}_c\ x_t\ r_t\ r) * C \mid D \Vdash (\sigma_t', b_t(r_t)) \mathrel{\overset{*}{\precsim}} (\sigma_s', b_s(r_s))) \end{array}}{C \mid D \Vdash (\sigma_t, \mathbf{call}(f, a_t) \gg\!= b_t) \mathrel{\overset{*}{\precsim}} (\sigma_s, \mathbf{call}(f, a_s) \gg\!= b_s)}$$

SIM-RET
$$\frac{Q\ \sigma_t\ \sigma_s\ r_t\ r_s}{(\sigma_t, \mathbf{return}\ r_t) \mathrel{\overset{*}{\precsim}} (\sigma_s, \mathbf{return}\ r_s)\ \{Q\}}$$

**Standard structural rules for weakest preconditions.**

SIM-MONO
$$\frac{(\sigma_t, b_t) \mathrel{\overset{*}{\precsim}} (\sigma_s, b_s)\ \{P\} \wedge \ulcorner P \vdash Q \urcorner}{(\sigma_t, b_t) \mathrel{\overset{*}{\precsim}} (\sigma_s, b_s)\ \{Q\}}$$

SIM-BIND
$$\frac{(\sigma_t, b_t) \mathrel{\overset{*}{\precsim}} (\sigma_s, b_s)\ \left\{\lambda \sigma_t'\ \sigma_s'\ r_t\ r_s.\, (\sigma_t', k_t(r_t)) \mathrel{\overset{*}{\precsim}} (\sigma_s', k_s(r_s))\ \{Q\}\right\}}{(\sigma_t, b_t \gg\!= k_s) \mathrel{\overset{*}{\precsim}} (\sigma_s, b_s \gg\!= k_t)\ \{Q\}}$$

SIM-FRAME
$$\frac{(\sigma_t, b_t) \mathrel{\overset{*}{\precsim}} (\sigma_s, b_s)\ \{Q\} * P}{(\sigma_t, b_t) \mathrel{\overset{*}{\precsim}} (\sigma_s, b_s)\ \{\lambda \sigma_t'\ \sigma_s'\ r_t\ r_s. (Q\ \sigma_t'\ \sigma_s'\ r_t\ r_s) * P\}}$$

SIM-SUPD
$$\frac{\Rrightarrow (\sigma_t, b_t) \mathrel{\overset{*}{\precsim}} (\sigma_s, b_s)\ \{\lambda \sigma_t'\ \sigma_s'\ r_t\ r_s.\, \Rrightarrow (Q\ \sigma_t'\ \sigma_s'\ r_t\ r_s)\}}{(\sigma_t, b_t) \mathrel{\overset{*}{\precsim}} (\sigma_s, b_s)\ \{Q\}}$$

Fig. 5. Selected rules for sepSim.

The first box contains standard simulation rules independent of separation logic. The top row shows rules for dual non-determinism. Following Sammler et al. [2023], we use notations $\dot{\forall} x \in X.$ for angelic non-determinism and $\dot{\exists} x \in X.$ for demonic non-determinism. These operators are best understood through the source-side rules ANGELIC-SRC and DEMONIC-SRC, where they correspond to universal and existential quantifiers respectively. When they are placed on the target side (ANGELIC-TGT, DEMONIC-TGT), the quantification is reversed. The second row shows standard simulation rules for states (**put** and **get**). The third row shows standard rules [Back and Wright 2012] for executing `assume`/`assert` commands. These commands are used to used to "lubricate" the refinement proof by discharging proof obligation to other, most appropriate layers. For example, when proving $T \sqsubseteq M \sqsubseteq S$, putting `assume`$(P)$ in $M$ means that the proof obligation $P$ is discharged to the upper (right) refinement, whereas the lower (left) refinement can rely on it (and vice versa for `assert`).

The second box presents separation-logic-related rules that are unique to this paper. In the first row are rules for executing `*assume`/`*assert` commands. Those are separation logic version of `assume`/`assert` commands, which only supported pure propositions. Rules for the former exactly follows the rules for the latter, except that it uses separating conjunction instead of pure conjunction and separating implication instead of pure implication. The rule for calling a function is the dual of the sepSimFun. The rule for returning from a function requires us to prove the postcondition, $Q$.

The third box comprises standard rules for weakest preconditions in separation logic. SIM-FRAME says that if one can prove the simulation and also has a separate $P$, one can still prove the simulation without $P$ and give it to the context at the end. SIM-MONO is self-explanatory.

The SIM-BIND rule is equivalent (assuming SIM-FRAME and SIM-MONO) to the following:

SIM-BIND-ALT

$$\frac{\exists R.\, (\sigma_t, b_t) \overset{*}{\lesssim} (\sigma_s, b_s)\ \{R\} \quad * \quad (\forall \sigma_t'\, \sigma_s'\, r_t\, r_s.\, R\, \sigma_t'\, \sigma_s'\, r_t\, r_s \twoheadrightarrow (\sigma_t', k_t(r_t)) \overset{*}{\lesssim} (\sigma_s', k_s(r_s))\ \{Q\})}{(\sigma_t, b_t \gg= k_s) \overset{*}{\lesssim} (\sigma_s, b_s \gg= k_t)\ \{Q\}}$$

saying that one can the proof obligation into two parts.

Finally, the SIM-SUPD rule. The rule allows us to freely insert $\dot{\Vdash\!\!\Rrightarrow}$. What this means is that when one is proving sepSim, one can freely "update" the separation logic conditions in the premise. While the rule is almost standard, there is one thing special here. Instead of the usual update modality $\dot{\Rrightarrow}$, we only support a slightly stronger modality, $\dot{\Vdash\!\!\Rrightarrow}$, which we call a **strong update modality**. For clarity, we henceforth call the original update modality a **weak** update modality.

As will be discussed in §7, there is a subtle counterexample showing that we cannot have weak update modality together with all the desiderata for sepSim; something should be weakened. After exploring various design spaces, the best solution we found is to strengthen the update modality (thus weakening sepSim).

The strong update modality still enjoys all the same structural rules as the weak one:

SUPD-MONO
$$\frac{P \vdash Q}{\dot{\Vdash\!\!\Rrightarrow} P \vdash \dot{\Vdash\!\!\Rrightarrow} Q}$$

SUPD-I
$$\frac{P}{\dot{\Vdash\!\!\Rrightarrow} P}$$

SUPD-E
$$\frac{\dot{\Vdash\!\!\Rrightarrow}\dot{\Vdash\!\!\Rrightarrow} P}{\dot{\Vdash\!\!\Rrightarrow} P}$$

SUPD-FRAME
$$\frac{\dot{\Vdash\!\!\Rrightarrow} P * Q}{\dot{\Vdash\!\!\Rrightarrow}(P * Q)}$$

SUPD-WUPD
$$\frac{\dot{\Vdash\!\!\Rrightarrow} P}{\dot{\Rrightarrow} P}$$

and the rule SUPD-WUPD shows that strong update modality is actually stronger. Also, the following key property SUPD-UPD allows one to update owned resources.

SUPD-UPD
$$\frac{a \rightsquigarrow b}{\boxed{a} \vdash \dot{\Vdash\!\!\Rrightarrow} \boxed{b}}$$

WUPD-UPD-NONDET
$$\frac{a \overset{set}{\rightsquigarrow} B}{\boxed{a} \vdash \dot{\Rrightarrow} \exists b \in B.\, \boxed{B}}$$

The only rule that gets invalidated with $\dot{\Vdash\!\!\Rrightarrow}$, compared to $\dot{\Rrightarrow}$, is the "non-deterministic" update rule, WUPD-UPD-NONDET, shown on the right. The definition of $\overset{set}{\rightsquigarrow}$ is given in §5.1. To our knowledge, in Iris, this rule is used only for allocating ghost names (including those for invariants [Jung et al. 2016]). CCR is in a very different setting, and no proof in CCR uses this rule.

## 5 FORMAL MODEL OF CCR 2.0

In this section, we discuss formal models of CCR 2.0.

### 5.1 Strong Update Modality

We define the strong update modality as follows (weak update modality presented for comparison):

$$\dot{\Rrightarrow} P \triangleq \lambda\, a.\, \forall c.\, \checkmark(a \cdot c) \Rightarrow \exists\, b.\, \checkmark(b \cdot c) \wedge P\, b \qquad a \rightsquigarrow b \triangleq \forall c.\, \checkmark(a \cdot c) \Rightarrow \checkmark(b \cdot c)$$

$$\dot{\Vdash\!\!\Rrightarrow} P \triangleq \lambda\, a.\, \exists b.\, a \rightsquigarrow b \wedge P\, b \qquad a \overset{set}{\rightsquigarrow} B \triangleq \forall c.\, \checkmark(a \cdot c) \Rightarrow \exists b \in B.\, \checkmark(b \cdot c)$$

$$
\begin{array}{lll}
\text{fundef}(E) & \triangleq \text{Any} \rightarrow \texttt{itree}\ E\ \text{Any} & \mathbb{T}|_{cond} \triangleq \text{if } cond \text{ holds, then } \mathbb{T} \text{ else } \emptyset \\
E_{\text{P}}(\mathbb{T}) & \triangleq \{\texttt{DemE}\} \uplus \{\texttt{AngE}\} \uplus \{\texttt{ObsE}\ f\ a \mid fn \in \textbf{String},\ a \in \text{Any}\}|_{\mathbb{T}=\text{Any}} & \\
E_{\text{M}}(\mathbb{T}) & \triangleq E_{\text{P}}(\mathbb{T}) \uplus \{\texttt{CallE}\ f\ a | f \in \textbf{String},\ a \in \text{Any}\}|_{\mathbb{T}=\text{Any}} \uplus \{\texttt{PutE}\ a \mid a \in \text{Any}\}|_{\mathbb{T}=\mathbf{1}} \uplus \{\texttt{GetE}\}|_{\mathbb{T}=\text{Any}} & \\
\text{PlainMod} & \triangleq \left\{(\texttt{funs}, \texttt{init_st}) \in (\textbf{String} \xrightarrow{\text{fin}} \text{fundef}(E_{\text{M}})) \times \text{Any}\right\} & \\
\hline
E_{\text{MC}}(\mathbb{T}) & \triangleq E_{\text{M}}(\mathbb{T}) \uplus \{{*}\texttt{AssumeE}\ P \mid P \in \textbf{sProp}\}|_{\mathbb{T}=\mathbf{1}} \uplus \{{*}\texttt{AssertE}\ P \mid P \in \textbf{sProp}\}|_{\mathbb{T}=\mathbf{1}} & \\
T, S \in \text{Mod} & \triangleq \left\{(\texttt{funs}, \texttt{init_st}, \texttt{init_res}) \in (\textbf{String} \xrightarrow{\text{fin}} \text{fundef}(E_{\text{M}})) \times \text{Any} \times \Sigma\right\} & \\
\Downarrow M & \triangleq \{(\lambda f\, a.\, \texttt{funs}_M(f)(a)[{*}\texttt{AssumeE} \mapsto \textbf{UB},\ {*}\texttt{AssertE} \mapsto \textbf{UB}], \texttt{init_st}_M)\} &
\end{array}
$$

Fig. 6. Definitions of plain module in CCR 1.0 (above) and enriched module in CCR 2.0 (below).

Here, ✓ is the standard validity condition of PCM. If we unfold the definition of $\rightsquigarrow$, the only difference between $\dot{\Rrightarrow}$ and $|\dot{\Rrightarrow}$ is the precedence of ∀ over ∃. The strong update modality only accepts a valid update under *arbitrary* context ($a \rightsquigarrow b$). The weak update modality allows one to peek into the context ($\forall c$. appears first) and then to pick an update ($\exists\, b$. appears later). Similarly, the definition of $\overset{set}{\rightsquigarrow}$ allows one to peek into the context and then pick an update. This is the reason why WUPD-UPD-NONDET holds for the weak update modality, but not for the strong update modality.

## 5.2 Module

The notion of a module in CCR 2.0 is presented in Fig. 6. It is based on the module in CCR 1.0, PlainMod, but is enriched with additional structures. First, PlainMod is based on Interaction Trees, and fundef(-)—a shorthand for `itree` s whose arguments and return types are Any—is central in our formalization. The primitive event, $E_{\text{P}}$, comprises demonic non-determinism, angelic non-determinism, and an observable event. Formally, $\dot{\exists} x \in X.\, k(x)$ is a shorthand for `trigger(DemE` `(`$\mathbb{T}$`))` $\gg\!= k$, and similarly for $\dot{\forall} x \in X.\, k(x)$ (triggering `AngE`) and *print* (triggering `ObsE`). The behavioral interpretation of demonic non-determinism is given as $\text{Beh}(\dot{\exists} x \in X.\, b(x)) \triangleq \bigcup_{x \in X} \text{Beh}(b(x))$, getting the union of all possible choices. Dually, we have $\text{Beh}(\dot{\forall} x \in X.\, b(x)) \triangleq \bigcap_{x \in X} \text{Beh}(b(x))$. Full formal definition of Beh is omitted here, and we refer interested readers to [Song et al. 2023].

With these, **assume** is simply defined as: $\textbf{assume}(P) \triangleq \textbf{if}\ P\ \textbf{then}\ \textbf{ret}(())\ \textbf{else}\ \dot{\forall}_ \in \emptyset.\ \textbf{ret}(())$ and dually for **assert** [Back and Wright 2012; Koenig 2020]. **UB** is defined as **assume**(⊥).

Then, $E_{\text{M}}$ (events for PlainMod) adds three more events: `CallE`, `PutE`, and `GetE`. Similar to before, each of these corresponds to a function call in the module system, **put**, and **get**, respectively. Then, a module comprises `funs`, a finite partial function from a function name to its semantics (given in ITrees), and `init_st`, an initial state for the module.

The Mod in CCR 2.0 uses $E_{\text{MC}}$ which extends the PlainMod with two additional events, `*AssumeE`, and `*AssertE`. Similarly, ***assume** is a shorthand for triggering `*AssumeE` and ***assert** for `*AssertE`. Mod also carries an initial resource, `init_res`, for establishing the module's invariant. It is a minor change from CCR 1.0: previously, the initial resource was specified in the wrapper, but it is more natural to be part of the module.

The user will almost always be working on Mod. However, in order to piggyback on the definition of contextual refinement defined for PlainMod, we maintain some connection between these. Our wrapper takes a Mod (written by the user) and returns PlainMod. Also, there is a trivial embedding, ⇑, from PlainMod to Mod (we use it implicitly). Finally, we define ⇓ operator that maps PlainMod into Mod and we will explain this operator in Theorem 5, the only place where it is used.

## 5.3 Definition of the Wrapper

We now define the wrapper in CCR 2.0, highlighting key changes from CCR 1.0.

For each function semantics, the wrapper (i) applies wrapF (putting conditions at the beginning and the end of the function), which internally calls wrapC (putting conditions before and after function calls), and then (ii) applies interpConds. Also, the wrapper packs the initial state and

```
⌊M⌋_C ≜ (λf ∈ dom(funs_M). interpConds(wrapF(C, C(f), funs_M(f))), (init_st_M, init_res_M))

wrapF(C, (X, P, Q), s ∈ fundef(E_MC)) ∈ fundef(E_MC) ≜ λa_t.
  ∀̇x ∈ X. a_s ← *assumeAndAbstr(P(x), a_t);
  r_s ← s(a_s)[CallE f a_s' ↦ wrapC(C(f), f, a_s'),
               PutE σ'     ↦ (_, μ) ← get; put(σ', μ); ret(()),
               GetE        ↦ (σ, _) ← get; ret(σ)];
  r_t ← *assertAndConcr(Q(x), r_s); ret(r_t)

wrapC((X, P, Q), f, a_s) ≜
  ∃̇x ∈ X. a_t ← *assertAndConcr(P(x), a_s);
  r_t ← call f a_t;
  r_s ← *assumeAndAbstr(Q(x), r_t); ret(r_s)

*assumeAndAbstr(PQ ∈ Any → Any → sProp, xr_t) ≜ ∀̇xr_s ∈ Any. *assume(PQ xr_s xr_t); ret(xr_s)
*assertAndConcr(PQ ∈ Any → Any → sProp, xr_s) ≜ ∃̇xr_t ∈ Any. *assert(PQ xr_s xr_t); ret(xr_t)

interpConds(s ∈ fundef(E_MC)) ∈ fundef(E_M) ≜ λa_t.
  let (_, r_t) := s(a_t)[*AssumeE(R ∈ sProp) ↦ λρ. interp*assume(R, ρ)
                         *AssertE(R ∈ sProp) ↦ λρ. interp*assert(R, ρ)](ε) in ret(r_t)

interp*assume(R ∈ sProp, ρ) ≜
  ∀̇γ ∈ Σ. assume(R(γ))
  (_, μ) ← get; assume(✓(μ · ρ · γ));
  ret(ρ · γ, ())

interp*assert(R ∈ sProp, ρ) ≜
  ∃̇γ ∈ Σ. assert(R(γ))
  (σ, μ) ← get; ∃̇(ρ', μ') ∈ Σ. assert((ρ · μ) ⇝ (ρ' · μ' · γ));
  put(σ, μ'); ret(ρ', ())
```

Fig. 7. Definition of the wrapper.

the initial resource into a pair, and accesses of state via PutE and GetE are adjusted accordingly in wrapF. The notation $[E_0 \mapsto (\lambda\sigma. H_0), E_1 \mapsto (\lambda\sigma. H_1), \ldots](\sigma : S)$ stands for interp_state in ITrees library, which interprets events $E_0, E_1, \ldots$ with corresponding handlers $H_0, H_1, \ldots$, where each handler is a state monad (it gets $\sigma$ and should return the changed state paired with its own return value) and $\sigma$ is the initial state.

The first phase (i) translates between the source and the target value. The *assumeAndAbstr() takes a condition and a target value and returns a source value satisfying the condition. Dually, *assertAndConcr() takes a condition and a source value and returns a target value satisfying the condition. For example, when proving $T \sqsubseteq \lfloor M \rfloor_C$, this allows the body of $M$ to work on its own arguments and return values ($a_s$ and $r_s$), as in the third line of wrapF ($r_s \leftarrow s(a_s)$), while the physical arguments and return values remain the same with the target ($a_t$ and $r_t$), an inherent requirement of contextual refinement. For more detail, since this functionality itself is not the contribution of this paper, we refer interested readers to [Sammler et al. 2023; Song et al. 2023].

The second phase (ii) now fully concretizes *assume/*assert, where their definitions are given in interp*assume/interp*assert. And here comes our key difference with CCR 1.0: CCR 1.0's model was centered around $\dot{\Rrightarrow}$, while the new model is centered around $|\!\Rrightarrow$. We concern three resources here: $\mu$ for module-private resources (which is used for $\mathcal{I}$), $\rho$ for function-scoped resources, and $\gamma$ for the resource that matches the condition. The module begins with empty $\rho$ ($\varepsilon$) and whenever *assume a proposition, it adds the corresponding resource ($\rho$).

A more interesting part is *assert: the condition $(\rho \cdot \mu) \rightsquigarrow (\rho' \cdot \mu' \cdot \gamma)$ allows one to update the resource only to the extent of SUPD-UPD; it does not justify WUPD-UPD-NONDET. CCR 1.0's model was very different from that it (angelically) picked a "context" resource at the beginning of *assume, following the style of WUPD-UPD-NONDET. After the update condition is checked, it updates $\rho$ and $\mu$ correspondingly, but throws away $\gamma$.

Finally, although the *model* checks validity at *assume and updates resources at *assert only, sepSim allows users to get validity and update resources at any point.

## 5.4 Adequacy

The adequacy result of CCR is stated with the following global wrapper elimination theorem:

$$\begin{array}{lll} \text{String} \ni f & \text{List(Val)} \ni \vec{v} & \text{Val} \to \text{Cmd} \ni b \\ \text{Val} \ni v & \triangleq & \texttt{int32}(i) \mid \texttt{long}(i) \mid \texttt{ptr}(p \in Loc \times \mathbb{Z}) \mid \not{z} \\ \text{Cmd} \ni c & \stackrel{\text{coind}}{=} & \texttt{calloc}(v) \mid *v := v \mid *v \mid \texttt{free}(v) \mid \textbf{call}(\vec{v}) \mid \texttt{print}(v) \mid x \leftarrow c; b \mid \textbf{return}(v) \mid \ldots \end{array}$$

Fig. 8. Syntax of the memLang.

**Theorem 5** (Wrapper Elimination Theorem (WET))**.** *For a resource algebra $\Sigma$, set of conditions $C$, a set of modules $M_1, \ldots, M_n$, an initial resource $\gamma_{\texttt{main}}$ satisfying the precondition of* `main`*, if the initial resources are consistent* (i.e., $\checkmark(\gamma_{\texttt{main}} \cdot \texttt{init_res}_{M_1} \cdot \ldots \cdot \texttt{init_res}_{M_n})$)*, we have the following:*

$$\lfloor S_1 \rfloor_C \oplus \ldots \oplus \lfloor S_1 \rfloor_C \qquad \sqsubseteq_{closed} \qquad \Downarrow S_1 \oplus \ldots \oplus \Downarrow S_n$$

## 6 RELATIONAL SEPARATION LOGIC ON TOP OF CCR 2.0

This section elaborates on the relational separation logic discussed in §2.2.

### 6.1 A Tiny Language, memLang, and Its Program Logic, memLangSim

To this end, we first define a tiny language with memory, called memLang, shown in Fig. 8. memLang is a language defined in a mixed-embedding style [Chlipala 2021] using Interaction Trees [Xia et al. 2019]. It has values and a memory model inspired by CompCert [Leroy 2006] and $\not{z}$ is an undef value. It has usual memory operations, function call, system call (*e.g.*, `print`), and return. The $\vec{v}$ notation denotes a list of values. A command can be sequenced via the bind operator $x \leftarrow c; b$ where the continuation $b$ can use the result of the former computation (bound to $x$ in the metalanguage). Its semantics is standard: for instance, the following pair of programs both prints 5 and returns a fresh pointer with size `40`.

$$\{\top\} \begin{array}{l} p \leftarrow \texttt{calloc}(8);\ *p := 5;\ x \leftarrow *p;\ \texttt{print}(x) \\ q \leftarrow \texttt{calloc}(40);\ \textbf{return}(q) \end{array} \quad \leq \quad \begin{array}{l} \texttt{print}(5) \\ q \leftarrow \texttt{calloc}(40);\ \textbf{return}(q) \end{array} \{v_t\, v_s.\, \mathcal{V}\, v_t\, v_s\}$$

We can prove this Hoare quadruple using memLangSim, a program logic for memLang. In memLangSim, there are *both* target points-to ($\mapsto^{\text{tgt}}$) and source points-to ($\mapsto^{\text{src}}$), each representing ownership for target and source memory. To be concrete, $p_t \mapsto^{\text{tgt}} \vec{v}_t$ means that the location from $p_t$ to $p_t + |\vec{v}_t|$ contains values $\vec{v}_t$. The rule MSIM-FOCUS-TARGET allows us to execute a single step in the target with "target unary rules" (second box in Fig. 9) and similarly for the source.

Now, in verifying the above example, we first apply TARGET-ALLOC (rpt $x$ $n$ is a list of length $n$ with all elements as $x$), TARGET-STORE, and TARGET-LOAD—along with MSIM-FOCUS-TARGET—and end up with:

$$\{\top\}\, \texttt{print}(5);\ q \leftarrow \texttt{calloc}(40);\ \textbf{return}(q) \leq \texttt{print}(5);\ q \leftarrow \texttt{calloc}(40);\ \textbf{return}(q)\ \{v_t\, v_s.\, \mathcal{V}\, v_t\, v_s\}$$

Now, we have the same commands on both sides which can be verified with "relational rules" (third box in Fig. 9). Two values are in value relation if they are identical scalar values (VREL-SCALAR-INTRO) or they originate from a lock-step allocation (BOTH-ALLOC). The fact that two values are related can be duplicated (VREL-DUP). We override the notation $\mathcal{V}$ for the list of values in BOTH-CALL. With these, the rest of the proof is a simple application of BOTH-PRINT, BOTH-ALLOC, and BOTH-RETURN.

Now, let us get back to a bird-eye view and see how memLangSim connects to CCR 2.0. In memLang, all the conditions on memory should be stated via ghost resources ($\mapsto^{\text{tgt}}$, $\mapsto^{\text{src}}$) and we intentionally hide physical memory from the user. To this end, we define a set of conditions VCond (and VConds respectively) that are parameterized over values but not memories. For $c \in$ VCond, we have trivial embedding $\hat{c} \in$ Cond: from $(\lambda x\, v_s\, v_t.\, P\, x\, v_s\, v_t)$ to $(\lambda x\, (\mathfrak{m}_s, v_s)\, (\mathfrak{m}_t, v_t).\, P\, x\, v_s\, v_t)$.

**Theorem 6** (Adequacy of memLangSim)**.** *For a pair of modules $T$ and $S$ written in* memLang *and $C \in$ VConds, if for every function name $f$ and its body $b_t$ in $T$, $b_s$ in $S$, and its condition $c$ in $C$ and*

**Structural Rules**

MSIM-FRAME
$$\frac{\{P\}\ c_t \le c_s\ \{\Phi\}}{\{P * R\}\ c_t \le c_s\ \{v_t\, v_s.\ \Phi\ v_t\ v_s * R\}}$$

MSIM-BIND
$$\frac{\{P\}\ c_t \le c_s\ \{v_t\, v_s.\ \Psi\ v_t\ v_s\} \qquad \forall v_t\, v_s.\ \{\Psi\ v_t\ v_s\}\ b_t(v_t) \le b_s(v_s)\ \{\Phi\}}{\{P\}\ x_t \leftarrow c_t;\ b_t \le x_s \leftarrow c_s;\ b_s\ \{\Phi\}}$$

MSIM-FOCUS-TARGET
$$\frac{\{P\}\ c_s\ \{v_s.\ \Psi\ v_s\}^{\mathrm{tgt}} \qquad \forall v_s.\ \{\Psi\ v_s\}\ c_t \le b_s(v_s)\ \{\Phi\}}{\{P\}\ c_t \le x_s \leftarrow c_s;\ b_s\ \{\Phi\}}$$

MSIM-FOCUS-SOURCE
$$\frac{\{P\}\ c_s\ \{v_s.\ \Psi\ v_s\}^{\mathrm{src}} \qquad \forall v_s.\ \{\Psi\ v_s\}\ c_t \le b_s(v_s)\ \{\Phi\}}{\{P\}\ c_t \le x_s \leftarrow c_s;\ b_s\ \{\Phi\}}$$

**Unary Rules**

$$p_t \mapsto^{\mathrm{tgt}} (\vec{v}_t +\!\!+\ \vec{w}_t) \equiv p_t \mapsto^{\mathrm{tgt}} \vec{v}_t * (p_t + |\vec{v}_t|) \mapsto^{\mathrm{tgt}} \vec{w}_t \qquad p_s \mapsto^{\mathrm{src}} (\vec{v}_s +\!\!+\ \vec{w}_s) \equiv p_s \mapsto^{\mathrm{src}} \vec{v}_s * (p_s + |\vec{v}_s|) \mapsto^{\mathrm{src}} \vec{w}_s$$

TARGET-ALLOC
$$\{\top\}\ \mathtt{calloc}(\mathtt{long}(i))\ \{p_t.\ p_t \mapsto^{\mathrm{tgt}} (\mathrm{rpt}\ 0\ (i/8))\}^{\mathrm{tgt}}$$

SOURCE-ALLOC
$$\{\top\}\ \mathtt{calloc}(\mathtt{long}(i))\ \{p_s.\ p_s \mapsto^{\mathrm{src}} (\mathrm{rpt}\ 0\ (i/8))\}^{\mathrm{src}}$$

TARGET-LOAD
$$\{p_t \mapsto^{\mathrm{tgt}} [v_t]\}\ {*}p_t\ \{v_t'.\ v_t' = v_t * p_t \mapsto^{\mathrm{tgt}} [v_t]\}^{\mathrm{tgt}}$$

SOURCE-LOAD
$$\{p_s \mapsto^{\mathrm{src}} [v_s]\}\ {*}p_s\ \{v_s'.\ v_s' = v_s * p_s \mapsto^{\mathrm{src}} [v_s]\}^{\mathrm{src}}$$

TARGET-STORE
$$\{p_t \mapsto^{\mathrm{tgt}} [_]\}\ {*}p_t := v_t\ \{p_t \mapsto^{\mathrm{tgt}} [v_t]\}^{\mathrm{tgt}}$$

SOURCE-STORE
$$\{p_s \mapsto^{\mathrm{src}} [_]\}\ {*}p_s := v_s\ \{p_s \mapsto^{\mathrm{src}} [v_s]\}^{\mathrm{src}}$$

**Relational Rules**

VREL-SCALAR-INTRO
$$\frac{v = \mathtt{int32}(i) \lor v = \mathtt{long}(i) \lor v = \text{↯}}{\mathcal{V}\ v\ v}$$

VREL-DUP
$$\frac{\mathcal{V}\ v_t\ v_s}{\mathcal{V}\ v_t\ v_s * \mathcal{V}\ v_t\ v_s}$$

BOTH-RETURN
$$\{\Phi\ v_t\ v_s * \mathcal{V}\ v_t\ v_s\}\ \mathbf{return}(v_t) \le \mathbf{return}(v_s)\ \{\Phi\}$$

BOTH-ALLOC
$$\{\mathcal{V}\ v_t\ v_s\}\ \mathtt{calloc}(v_t) \le \mathtt{calloc}(v_s)\ \{v_t'\, v_s'.\ \mathcal{V}\ v_t'\ v_s'\}$$

BOTH-LOAD
$$\{\mathcal{V}\ v_t\ v_s\}\ {*}v_t \le {*}v_s\ \{v_t'\, v_s'.\ \mathcal{V}\ v_t'\ v_s'\}$$

BOTH-STORE
$$\{\mathcal{V}\ v_t\ v_s * \mathcal{V}\ v_t'\ v_s'\}\ {*}v_t := v_t' \le {*}v_s := v_s'\ \{\top\}$$

BOTH-PRINT
$$\{\top\}\ \mathtt{print}(\mathtt{long}(i)) \le \mathtt{print}(\mathtt{long}(i))\ \{\top\}$$

BOTH-CALL
$$\frac{C(f) = c}{\{\mathcal{V}\ \vec{v}_t\ \vec{v}_s * P_c\ x\ v_s\ v_t * \mathcal{I}\ \sigma_t\ \sigma_s\}\ (\sigma_t, \mathbf{call}(f, \vec{v}_t)) \le (\sigma_s, \mathbf{call}(f, \vec{v}_s))\ \{v_t'\, v_s'.\ \mathcal{V}\ v_t'\ v_s' * Q_c\ x\ v_s\ v_t * \mathcal{I}\ \sigma_t\ \sigma_s\}}$$

Fig. 9. Selected rules for memLangSim.

*every argument* $\vec{v}_t$ *and* $\vec{v}_s$ *we have:*

$$\{\mathcal{V}\ \vec{v}_t\ \vec{v}_s * P_c\ x\ v_s\ v_t * \mathcal{I}\ \sigma_t\ \sigma_s\}\ (\sigma_t, b_t(\vec{v}_t)) \le (\sigma_s, b_s(\vec{v}_s))\ \{\sigma_t\ \sigma_s\ v_t\ v_s.\ \mathcal{V}\ v_t\ v_s * Q_c\ x\ v_s\ v_t * \mathcal{I}\ \sigma_t\ \sigma_s\}$$

*then we have:* $\mathcal{J} * \hat{C} \vDash T \sqsubseteq S$.

### 6.2 Instantiating memLang and memLangSim on Top of CCR 2.0

In this section, we show how memLang and memLangSim connect to CCR 2.0 in more detail. It is worth emphasizing memLangSim is really just a shallow layer on top of CCR 2.0. This is due to the use of ITrees in our formalization: ITrees and its event-based semantics allow adding layers while reusing lots of the rules and theorems.

Specifically, a module in memLang is simply a Mod in Fig. 6 but with event $E_{\mathrm{M}} + E_{\mathrm{mem}}$, enriched from $E_{\mathrm{M}}$. This formulation allows memLang to inherit operators and their rules defined in $E_{\mathrm{M}}$ for free, including `assume`, `assert`, **put**, **get**, and **call**. Then, $E_{\mathrm{mem}}$ adds four memLang-specific operations: calloc, load, store, and free. Given a memLang code $c$ and an initial memory $\mathfrak{m}_0$, the interpretation operator $[\![c]\!]^{\mathsf{mem}}_{\mathfrak{m}_0}$ interprets $E_{\mathsf{mem}}$ away and returns an interaction tree of type `itree` $E_{\mathrm{M}}$ Val and a resulting memory, $\mathfrak{m}_1$. With these, we have the following key definitions:

**Definition 7.** The relational separation logic condition, $\mathcal{J}$, is defined for any $f \in$ **String** as follows:

$$\mathcal{J}(f) \triangleq \{\mathbf{1},\ \lambda_\ (\mathfrak{m}_s, \vec{v}_s)\ (\mathfrak{m}_t, \vec{v}_t).\ \mathrm{inv}\ \mathfrak{m}_t\ \mathfrak{m}_s * \mathcal{V}\ \vec{v}_t\ \vec{v}_s,\ \ \lambda_\ (\mathfrak{m}_s, v_s)\ (\mathfrak{m}_t, v_t).\ \mathrm{inv}\ \mathfrak{m}_t\ \mathfrak{m}_s * \mathcal{V}\ v_t\ v_s\}$$

**Definition 8.** memLangSim is defined as follows:

$$\{P\}\ (\sigma_t, c_t) \leq (\sigma_s, c_s)\ \{Q\} \triangleq P \twoheadrightarrow \forall \mathfrak{m}_t\ \mathfrak{m}_s.\ \mathrm{inv}\ \mathfrak{m}_t\ \mathfrak{m}_s \twoheadrightarrow (\varepsilon \mid \mathcal{J} \Vdash (\sigma_t, [\![c_t]\!]^{\mathsf{mem}}_{\mathfrak{m}_t}) \stackrel{*}{\lesssim} (\sigma_s, [\![c_s]\!]^{\mathsf{mem}}_{\mathfrak{m}_s})\{Q\})$$

Here, inv maintains necessary ghost states to connect ghost resources, $\mapsto^{\mathrm{src}}$, $\mapsto^{\mathrm{tgt}}$, and $\mathcal{V}$, with actual physical memory. Note that $\mathcal{J}(f)$ specifies conditions not only on values but also on memories. Finally, we have the following interchangeability theorem.

**Theorem 9** (Interchangeability). *For $j, i \in$ Cond and $c, d \in$ VCond, if we have the following:*

$$\forall x\, x'\, \mathfrak{m}_m\ \mathfrak{m}_t\ v_m\ v'_m\ v_t.\quad (P_j\ x\ (\mathfrak{m}_m, v_m)\ (\mathfrak{m}_t, v_t) * P_c\ x'\ v'_m\ v_t) \vdash \ulcorner v_m = v_t \wedge v'_m = v_t \urcorner$$
$$\forall x\, x'\, \mathfrak{m}_m\ \mathfrak{m}_t\ v_m\ v'_m\ v_t.\quad (Q_j\ x\ (\mathfrak{m}_m, v_m)\ (\mathfrak{m}_t, v_t) * Q_c\ x'\ v'_m\ v_t) \vdash \ulcorner v_m = v_t \wedge v'_m = v_t \urcorner$$

*then we have $(j * c) \ltimes (i * d) \equiv (j \ltimes i) * (c \ltimes d)$.*

The rationale for this theorem is twofold: First, $c$ and $d$ should be unary (thus $v'_m = v_t$). Second, changes in the relational condition ($j$) should not affect the unary condition (thus $v_m = v_t$, though $\mathfrak{m}_m$ and $\mathfrak{m}_t$ may differ). Note that this theorem can be repeatedly applied to formulae comprising more than two clauses: *e.g.*, $(j * c) \ltimes (i * d) \ltimes (k * e) \equiv (j \ltimes i \ltimes k) * (c \ltimes d \ltimes e)$.

### 6.3 Evaluation

Implementing the ideas in the paper was non-trivial due to an inherent complexity of coinduction and separation logic models. On top of CCR 1.0, we add (excluding modifications) 7,596 SLOC of Rocq code (335 for definitions, 6,154 for proofs, and 1,107 for examples) to build CCR 2.0 (counted with `coqwc`). Additionally, we add 3,023 SLOC for relational separation logic and 1,031 SLOC for examples using it. In total, our development adds (excluding modifications) 11,650 SLOC of Rocq.

## 7 COUNTEREXAMPLES FOR I AND II IN CCR

In this section, we first discuss the two counterexamples mentioned in §4.1 and §4.2, respectively. These counterexamples rely on the generality of the CCR in that it supports arbitrary (non-step-indexed) resource algebra and arbitrary dual non-determinism, and we make contrived examples with those. Then, we present an advanced feature of CCR: changing arguments/return values.

### 7.1 Counterexamples

***The first example.*** The first counterexample is as follows. We use the following instances:

$$T = (\mathbf{def}\ f_t(x)\ \{\mathbf{skip}\}) \qquad S = (\mathbf{def}\ f_s(x)\ \{\mathbf{skip}\})$$
$$C(f) = (\mathbf{1}, \top, \lambda__.\boxed{a}) \qquad D(f) = (\mathbf{1}, \top, \bot) \qquad FR(f) = (\mathbf{1}, \lambda__.\boxed{a}, \top)$$

Here, $a$ can be an arbitrary non-unit resource. In other words, $\top \nvdash \boxed{a}$.

First, note that the first refinement, $\lfloor T \rfloor_C \sqsubseteq \lfloor S \rfloor_D$, is not provable using sepSim. There is no way one can prove $\bot$ at the end of the source. *However*, this refinement is actually provable in the *model* of CCR, exploiting the ill-formedness of the target.

The postcondition in the target (roughly) unfolds to `assert`($\exists q.\, p \rightsquigarrow q \wedge [\![\boxed{a}]\!]\ q$) for $p$ chosen by us, satisfying $[\![\top]\!]\ p$. If we pick $p$ as an empty resource, $\varepsilon$, the `assert` statement reduces to `assert`($\bot$) since one cannot *forge* a out of nothing. Thus, executing the target postcondition yields $\bot$, concluding the proof without needing to execute the source postcondition.

$T_{\mathrm{BS}}$
```
def bitset: long = 0

def set(key: long) ≡
  bitset := (bitset & (1 << k))

def get(key: long): long ≡
  ret (bitset | (1 << k))
```

$M_{\mathrm{BS}}$
```
def n: ℕ = 0

def set(k: ℕ): 1 ≡
  n := (Nat.setbit n k)

def get(k: ℕ): 𝔹 ≡
  ret (Nat.testbit n k)
```

$S_{\mathrm{BS}}$
```
private map ∈ Flag → 𝔹 := λk.⊥

def set(k: Flag): 1 ≡
  map := map[k ← ⊤]

def get(k: Flag): 𝔹 ≡
  ret map(k)
```

$C_{\mathrm{BS}}$

$$C(\mathsf{set}) \triangleq (\mathbf{1}, \{\lambda a_m\, a_t.\ \ulcorner a_m = l2n(a_t) \urcorner \wedge \ulcorner a_m \in [0, 64) \urcorner\}, \{\lambda r_m\, r_t.\ \ulcorner r_m = r_t = () \urcorner\})$$
$$C(\mathsf{get}) \triangleq (\mathbf{1}, \{\lambda a_m\, a_t.\ \ulcorner a_m = l2n(a_t) \urcorner \wedge \ulcorner a_m \in [0, 64) \urcorner\}, \{\lambda r_m\, r_t.\ \ulcorner r_m = l2n(r_t) \urcorner\})$$

$D_{\mathrm{BS}}$

$$D(\mathsf{set}) \triangleq (\mathbf{1}, \{\lambda a_s\, a_m.\ \ulcorner f2n(a_s) = a_m \urcorner\}, \{\lambda r_s\, r_m.\ \ulcorner r_s = r_m = () \urcorner\})$$
$$D(\mathsf{get}) \triangleq (\mathbf{1}, \{\lambda a_s\, a_m.\ \ulcorner f2n(a_s) = a_m \urcorner\}, \{\lambda r_s\, r_m.\ \ulcorner r_s = r_m \urcorner\})$$

Fig. 10. An implementation $T_{\mathrm{BS}}$, its intermediate abstraction $M_{\mathrm{BS}}$, and its final abstraction $S_{\mathrm{BS}}$ (above). Conditions used in the lower refinement $C$, and conditions used in the upper refinement $D$ (below).

Note that $\lfloor T \rfloor_C$ is an ill-formed module. No implementation (think of modules written in usual programming languages, like C, without conditions) can refine $\lfloor T \rfloor_C$ (doing so would require proving $\bot$). Thus, from the user's perspective, $\lfloor T \rfloor_C$ can never appear in the verification scenario.

Next, consider the second refinement. The target now becomes a sensible module due to $\boxed{a}$ in its precondition. Specifically, its postcondition now reduces to `assert`$(\top)$—any $p$ satisfying the precondition also satisfies the postcondition. This means we cannot establish the second refinement using the same trick as we did in the first refinement. One needs to genuinely prove $\bot$ in the source postcondition, which is impossible.

Thus, since the first refinement is provable while the second refinement is not, this constitutes a counterexample for VFRAME?.

***The second example.*** For space reasons, we defer detailed construction of the second example to the appendix and discuss high-level points here.

The second example shows that if sepSim supported $\dot{\Rrightarrow}$ instead of $|\dot{\Rrightarrow}$, there is a contradiction. Both the counterexample and also the argument for that being a counterexample are much more involved and interesting here. The counterexample exploits (i) sepSim's support for full dual non-determinism, (ii) the power of weak update modality (WUPD-UPD-NONDET), and (iii) a subtle, unpresented (for the sake of space) feature of sepSim, to create a "race condition" between upper and lower refinements, updating resources in conflicting ways.

### 7.2 Advanced Feature: Changing Arguments and Return Values

In this section, we show a minimal example that uses a feature of CCR that we haven't showed so far in previous examples: changing arguments and return values. Using this feature further clarifies the difference between $\ltimes$ and $*$. We have intentionally minimized the example so that it does not use any separation logic conditions, neither unary nor relational.

The implementation module $T_{\mathrm{BS}}$ implements a bitset using `long` type. The middle abstraction $M_{\mathrm{BS}}$ abstracts it to a natural number and uses operators defined in `Nat` library of Rocq. This refinement relies on conditions that relate the natural number with `long`; for instance, consider the condition for `get` in $C$. Using Hoare-Quadruple-like notation, the condition in Fig. 13 is expressed like this:

$$\forall a_m\, a_t.\ \{\ulcorner a_m = l2n(a_t) \urcorner \wedge \ulcorner a_m \in [0, 64) \urcorner\}\ (r_t \leftarrow \mathsf{get}(a_t)) \le (r_m \leftarrow \mathsf{get}(a_m))\ \{\ulcorner r_m = r_t = () \urcorner\}$$

where $l2n(-)$ converts `long` into $\mathbb{N}$. Since $\mathbb{N}$ has strictly more elements than `long`, conditions like $\ulcorner a_m = l2n(a_t)\urcorner$ is essential, requiring the client to pass only the natural numbers within the range.

Finally, the source module $S_{\text{BS}}$ further abstracts $\mathbb{N}$ into $Flag$, which is a custom algebraic data type defined in Rocq. The datatype also has an embedding into $\mathbb{N}$, $f2n(-)$. The conditions for the upper refinement, $D$, are largely similar to $C$ except that it relates $\mathbb{N}$ and $Flag$.

Now, let us see how $\ltimes$ and $*$ behaves differently here.

$$C \ltimes D(\texttt{get}) = (\mathbf{1}, \{\lambda a_s\, a_t.\, \ulcorner f2n(a_s) = l2n(a_t)\urcorner\}, \quad \{\lambda r_s\, r_t.\, \ulcorner r_s = l2n(r_t)\urcorner\})$$
$$C * D(\texttt{get}) = (\mathbf{1}, \{\bot\}, \quad \{\bot\})$$

The resulting pre/postcondition in $C \ltimes D$ adequately relates the arguments and return values in $T_{\text{BS}}$ and $S_{\text{BS}}$, whereas conditions on $M_{\text{BS}}$ is disappeared. On the other hand, $C * D$ gives an ill-formed pre/postcondition that are equivalent to just $\bot$. To see this more clearly, the precondition of $C * D(\texttt{get})$ is computed as follows:

$$\{\lambda a_{sm}\, a_{mt}.\, \ulcorner a_{sm} = l2n(a_{mt})\urcorner \wedge \ulcorner f2n(a_{sm}) = a_{mt}\urcorner\}$$

As we can see, now the first variable, $a_{sm}$, stands for both the argument in $S_{\text{BS}}$ and $M_{\text{BS}}$ (and similarly for $a_{mt}$). That is, in the first clause ($a_{sm} = l2n(\ldots)$), it behaves as an argument in $M_{\text{BS}}$ and has type $\mathbb{N}$. In the second clause ($f2n(a_{sm}) = \ldots$), it behaves as an argument in $S_{\text{BS}}$ and has type $Flag$. This is contradictory, and thus the resulting condition is $\bot$. Similarly for the postcondition.

CCR 1.0 did not have an operator $\ltimes$: this is a newly defined one in CCR 2.0. As a side note, this example also shows Problem **I** again in the same sense as before: If we have $T'_{\text{BS}}$ that is the same with $T_{\text{BS}}$ but using `int32` instead of `long`, one cannot accommodate proof reuse in CCR 1.0.

## 8 RELATED WORK

Simuliris [Gäher et al. 2022] developed a relational separation logic that supports concurrency whereas sepSim and memLangSim do not. However, as discussed in §2.2, the adequacy theorem of Simuliris enforces every function to have trivial pre/postconditions. On the other hand, memLangSim supports arbitrary non-trivial conditions in its adequacy theorem (Theorem 6).

Parkinson and Summers [2011] first observed the connection between dual non-determinism and separation logic conditions and greatly influenced projects like Viper [Müller et al. 2016]. However, both Parkinson and Summers [2011] and Viper are unary logics and do not support refinement.

Recent increasing interest in the use of dual non-determinism in the refinement setting is pioneered by Koenig [2020]; Koenig and Shao [2020]. They use dual non-determinism to support relational reasoning similar to CCR 2.0. However, they support only pure conditions whereas CCR 2.0 supports separation logic conditions.

DimSum [Sammler et al. 2023] is inspired by CCR and shares numerous characteristics (especially, the use of wrapper and dual non-determinism), but have different goals: modeling multi-language semantics versus having a program verification technique that fuses different styles. The key benefit of DimSum over CCR is its support for arbitrary user-provided events (like assembly's **`jump`**) for *module interaction*, whereas CCR supports only the **`call`** event for module interaction. [7] DimSum's wrapper is designed to bridge the gap between different languages and events, and they use relational separation logic conditions akin to Simuliris in their wrapper. *However*, like Simuliris, DimSum's wrapper does not support user-defined non-trivial conditions whereas CCR does. Thus, DimSum's wrapper does not support verifications like Fig. 2 where conditions ($E_{\texttt{Map}}$) are built through step-wise refinement ($C_{\texttt{Map}}$ and $D_{\texttt{Map}}$).

[7] As seen in §6, one can have arbitrary module-internal events like $E_{\text{mem}}$ as long as they will get interpreted to $E_{\text{MC}}$.

## 9 LIMITATIONS AND FUTURE WORK

At the moment, CCR does not support any form of concurrency. Extending CCR with concurrency is an ongoing work in parallel, and we do not see any fundamental inconsistency between that and CCR 2.0. CCR does not support step-indexing; but, on the other hand, CCR establishes termination-sensitive refinement. It remains an interesting research question to see if CCR can be extended to support techniques based on step-indexing.

We are exploring two directions on top of CCR 2.0, using its new capability (**VC** × **USL** × **RSL**).

The first is the use of verified translators (similar to compilers, but designed for abstracting modules toward specifications) for semi-automatic verification. For instance, instead of directly verifying an implementation, one could first apply *constant folding* translation to the implementation and then begin verification on the resulting output. In CCR 1.0, this was not well-supported since there could be only one memory-abstracting translation, but is now supported well in CCR 2.0.

The second is complier verification in the presence of sophisticated protocols like ownership types. Here, the state-of-the-art is to provide an intensional memory model that understands ownership principles [Jung et al. 2019] and verify against such a model. We are investigating whether CCR 2.0's capability to specify function-wise specifications while using relational separation logic could give another more extensional formulation to justify ownership-based optimizations.

## ACKNOWLEDGMENTS

This material is based upon work supported by the National Science Foundation under Grant No. nnnnnnn and Grant No. mmmmmmm. Any opinions, findings, and conclusions or recommendations expressed in this material are those of the author and do not necessarily reflect the views of the National Science Foundation.

$$\frac{(P \vdash Q) \land \text{cur_sProps } a\ P}{\text{cur_sProps } a\ Q}\ \textsc{cur-sProps-mono} \qquad \frac{\text{cur_sProps } a\ (\forall x \in X.\, P\ x)}{\forall x \in X.\, \text{cur_sProps } a\ P\ x}\ \textsc{cur-sProps-forall} \qquad \frac{\text{cur_sProps } a\ (\exists x \in X.\, P\ x)}{\exists x \in X.\, \text{cur_sProps } a\ P\ x}\ \textsc{cur-sProps-exists}$$

$$\frac{\text{cur_sProps } a\ (\ulcorner P \urcorner * Q)}{P \land \text{cur_sProps } a\ Q}\ \textsc{cur-sProps-pure} \qquad \frac{\text{cur_sProps } a\ (\dot{\Rrightarrow} P)}{P}\ \textsc{cur-sProps-supd} \qquad \frac{\text{cur_sProps } a\ \boxed{a'}}{\checkmark\ a'}\ \textsc{cur-sProps-valid}$$

Fig. 11. Selected rules for cur-sProps.

## A KEY IDEAS FOR BRIDGING THE GAP BETWEEN THE MODEL OF CCR AND sepSim

While the interface we built is concise, nontrivial engineering was needed under the hood to make it work. In usual separation logic built on top of Iris base logic, regardless of whether it is unary or relational [?], the developers of the logic never need to do model-level proof. There, the developer of the logic (almost) never needs to do model-level proof. In CCR, we had a different goal from these (having benefits of both refinements and separation logic) and we are in a unique situation where we need to do model-level proof. CCR integrates separation logic conditions as part of an *operational semantics*, in the form of `*assume`/`*assert` instructions, and these instructions comprise resource manipulations. In sepSim, we need to connect this model to the logic where no previous work had done anything similar.

We briefly show two key ideas that could be useful in other contexts.

***On asynchrony.*** An astute reader might have noticed that, while the `*assume`/`*assert` keywords are only inserted at the interaction points, the user can freely update the separation logic conditions (achieved via `*assume`) using sim-supd, not just at the interaction points.

Consider proving the following simple case where we have `*assume`/`*assert` only on the source:

$$(\textbf{def } f_t(x) \triangleq \textbf{skip}) \sqsubseteq (\textbf{def } f_s(x) \triangleq P; \textbf{skip}; \textbf{skip}; Q)$$

In the *operational semantics*, the source program roughly interprets as follows: executing $P$ gives you a resource $p$ satisfying $[\![P]\!]\ p$, and executing Q roughly unfolds to $\texttt{assert}(\exists q.\, p \rightsquigarrow q \land [\![Q]\!]\ q)$. Now, using sepSim, suppose that the user has updated $P$ into $R$ (satisfied by $r$) in the middle of the execution; *i.e.*, between the two **skip**s. Then, this resource $r$ and the proposition $R$ never appears in the operational semantics, and sepSim should maintain some connection between this $R$, $r$ and $p$, so that we can execute `assert` at the end.

For maintaining such an invariant in a manageable way (when validating the rules in sepSim), we use an abstraction called cur-sProps defined simply as follows:

$$\text{cur_sProps } p\ R \triangleq \checkmark(p) \land \exists r.\, p \rightsquigarrow r \land [\![R]\!]\ r$$

Then, we always maintain cur_sProps $p$ $R$ where $p$ is the resource the source module owns in operational semantics, and $R$ is separation logic propositions that appear to the user.

This abstraction admits the rules shown in Fig. 11. Using these rules minimized our exposure to a direct model-level proof in developing sepSim.

***On handling separation logic conditions on the target side.*** The next challenge is about dealing with separation logic conditions on the target side. Consider the following simple refinement:

$$(\textbf{def } f_t(x) \triangleq R; b_t; S) \sqsubseteq (\textbf{def } f_s(x) \triangleq P * R; b_s; Q * S)$$

Here, what sepSim provides to the user is that: if you give $R$ to the target (when executing $R$), you can get $S$ back from the target (when executing S). However, while this gives an *illusion* that we are actually, physically giving (and getting back) the proposition $R$ to (and $S$ from) the target, that is

PEEKING-WAND
$$\frac{(P \mathrel{-\!\!*} Q) * {}^{R}|\!\dot{\Rrightarrow} P}{{}^{R}|\!\dot{\Rrightarrow} Q}$$

PEEKING-ANTI
$$\frac{(S \mathrel{-\!\!*} R) * {}^{R}|\!\dot{\Rrightarrow} P}{{}^{S}|\!\dot{\Rrightarrow} P}$$

PEEKING-TRUE
$${}^{\top}|\!\dot{\Rrightarrow} P \dashv\vdash |\!\dot{\Rrightarrow} P$$

PEEKING-JOIN
$$\frac{{}^{R}|\!\dot{\Rrightarrow} {}^{R}|\!\dot{\Rrightarrow} P}{{}^{R}|\!\dot{\Rrightarrow} P}$$

PEEKING-∗
$$\frac{{}^{R}|\!\dot{\Rrightarrow} P * {}^{R}|\!\dot{\Rrightarrow} Q}{{}^{R}|\!\dot{\Rrightarrow} (P * Q)}$$

PEEKING-I
$$\frac{(p \cdot r) \rightsquigarrow (q \cdot r)}{\boxed{p} \vdash {}^{\boxed{r}}|\!\dot{\Rrightarrow} \boxed{q}}$$

PEEKING-E
$$\frac{P * {}^{R*P}|\!\dot{\Rrightarrow} Q}{{}^{R}|\!\dot{\Rrightarrow} P * Q}$$

Fig. 12. Selected rules for peeking update modality.

*not* what is happening under the hood. Recall the definition of refinement: it simply computes the behavior of the source and target—*in a unary manner*—and compares them. The notion of behavior and refinement does not come with a notion of communication between the source and target that one can physically give and get back those resources.

Thus, in order to give such an illusion to the user, we employ a technique we call **freezing**. While a primitive form of the idea itself was already present in Song et al. [2023], it was specific to one example in their model-level proof. It is our contribution to generalize it to arbitrary propositions and give a logic-level interface (sepSim) to the user.

The technique works this way. At the beginning, in the source, we get $pr$ satisfying $[\![P * R]\!]\ pr$. And, at the end of the function, we need to execute `assert`$(\exists qs.\ pr \rightsquigarrow qs \wedge [\![Q * S]\!]\ qs)$. For this, we first split $pr$ into $p$ (satisfying $[\![P]\!]\ p$) and (satisfying $[\![R]\!]\ r$). Then, we initiate the execution of the target with $r$. However, we do *not* (cannot) physically give $r$ away to the target. Instead, the $r$ resource still resides on the source side, but we **freeze** it and hide it from the user, so that it does not get altered by the user's manipulations. Technically, we always maintain cur_sProps $(p \cdot r)$ $(\ldots * R)$ where ... is what appears in the user's proof buffer and $R$ is frozen and hidden from the user.

Now, at the end of the function, we get (i) $\exists s.\ r \rightsquigarrow s \wedge [\![S]\!]\ s$ from the target. Also, what user has proven with sepSim is reified into the form of cur_sProps $(p \cdot r)$ $(Q * \boxed{r})$, thus we get (ii) $\exists q.\ (p \cdot r) \rightsquigarrow (q \cdot r) \wedge [\![Q]\!]\ q$. Combining (ii) and (i), we can execute `assert`$(\exists qs.\ pr \rightsquigarrow qs \wedge [\![Q * S]\!]\ qs)$ in the source.

The key invariant we maintain here is that the source resource always contains the target resource (in the frozen form). The same applies not just to resources for pre/postconditions, but also for the resource for invariants: that is the rationale behind the condition (*ii*) in Theorem 2.

# B COUNTEREXAMPLE FOR THE MODEL OF CCR 1.0

## B.1 Setup: Strengthening sepSim with Peeking

The second counterexample is much more involved. We first set up some technical machinery.

***Exploiting frozen resources.*** Recall that we have validated sepSim using the **freezing** technique (Appendix A). This actually provides more capabilities to the user than what is presented in sepSim: the user can *update* the resource *knowing that* $r$ is present. In other words, even though we are not modifying the frozen $r$, just knowing its presence gives strictly more capability than knowing nothing. Formally, there is a resource algebra and its element, $\mathrm{r}$, such that the following holds:

$$\{p, q \mid p \rightsquigarrow q\} \subsetneq \{p, q \mid (p \cdot r) \rightsquigarrow (q \cdot r)\}$$

and this forms the basis for our counterexample.

***The peeking update modality.*** Such an idea is formalized at the logic level as follows. We define what we call the **peeking update modality** as follows:

$${}^{R}|\!\dot{\Rrightarrow} P \triangleq \forall F.\ (F \wedge R) \mathrel{-\!\!*} |\!\Rrightarrow ((F \wedge R) * P)$$

and similarly define ${}^{R}\dot{\Rrightarrow}$ using $\dot{\Rrightarrow}$. Here, $R$ captures the "frozen" resource and the universal quantification of $F$ ensures that the resource in $R$ is unchanged during the update, a common pattern in Iris [Sammler et al. 2021].

The rules for ${}^{R}|\dot{\Rrightarrow} P$ are shown in Fig. 12. While we have shown the rules for ${}^{R}|\dot{\Rrightarrow}$ in the figure, exactly the same rules with ${}^{R}\dot{\Rrightarrow}$ and $\dot{\Rrightarrow}$ (instead of ${}^{R}|\dot{\Rrightarrow}$ and $|\dot{\Rrightarrow}$) hold as well.

The rule PEEKING-WAND and PEEKING-ANTI says that it is monotone and antitone with its two arguments, respectively. PEEKING-TRUE says that if the frozen proposition is trivial, then it is equivalent to the underlying update modality. PEEKING-JOIN says that ${}^{R}|\dot{\Rrightarrow}$ satisfies a monadic structure (together with PEEKING-MONO), and PEEKING-$*$ says that ${}^{R}|\dot{\Rrightarrow}$ is a modality in separation logic in the sense of Krebbers et al. [2018].

The rule PEEKING-I and PEEKING-E are the main interesting parts. The former is like SUPD-UPD, but it allows "peeking" the resource $r$. The latter allows one to (partially) eliminate the ${}^{R}|\dot{\Rrightarrow}$ and make it closer to its underlying update modality. With these, the following rules are derived:

PEEKING-E-FULL
$$\frac{P * {}^{P}|\dot{\Rrightarrow} Q}{|\dot{\Rrightarrow} (P * Q)}$$

PEEKING-MONO
$$\frac{P \vdash Q}{{}^{R}|\dot{\Rrightarrow} P \vdash {}^{R}|\dot{\Rrightarrow} P}$$

PEEKING-E-FULL is derived by applying PEEKING-E, PEEKING-ANTI, and PEEKING-TRUE. PEEKING-MONO is derived by applying PEEKING-WAND.

sepSim$^{+}$***: Strengthening*** sepSim ***with Peeking.*** With this, we can strengthen sepSim into sepSim$^{+}$ that utilizes this additional capability. It carries the frozen resources in a freezer, $\boxed{\phantom{R}}$, as follows:

$$C; D \Vdash \boxed{R}\, (\sigma_t, b_t) \overset{*}{\gtrsim} (\sigma_s, b_s)\ \{Q\}$$

Also, sepSimFun is adjusted accordingly into sepSimFun$^{+}$:

$$\text{sepSimFun}^{+}\ C\, D\, \mathcal{I}\ f\ b_t\ b_s \triangleq \textbf{let}\ (c, d)\ :=\ (C(f), D(f))\ \textbf{in}\ \forall \sigma_t\ \sigma_s\ a.$$
$$\forall x_s \in \mathsf{X}_d.\, (\mathcal{I}\ \sigma_t\ \sigma_s) \twoheadrightarrow (\mathsf{P}_d\ x_s\ a) \twoheadrightarrow \exists x_t \in \mathsf{X}_c.\, (\mathsf{P}_c\ x_t\ a) * C; D \Vdash \boxed{(\mathsf{P}_c\ x_t\ a)}\, (\sigma_t, b_t(a)) \overset{*}{\gtrsim} (\sigma_s, b_s(a))\ \{\ldots\}$$

where it now carries the condition $(\mathsf{P}_c\ x_t\ a)$, given to the target, on its freezer. Finally, we now have a stronger version of SIM-SUPD as follows:

SIM-PEEKING-SUPD
$$\frac{{}^{R}|\dot{\Rrightarrow} \boxed{\mathrm{R}}\, (\sigma_t, b_t) \overset{*}{\gtrsim} (\sigma_s, b_s)\ \{Q\}}{\boxed{\mathrm{R}}\, (\sigma_t, b_t) \overset{*}{\gtrsim} (\sigma_s, b_s)\ \{Q\}}$$

This strengthening is sound in the model of CCR 2.0.

### B.2 Counterexample for Having $\dot{\Rrightarrow}$ in sepSim$^{+}$

Finally, we are ready to see the counterexample. We are going to prove a contradiction assuming sepSim$^{+}_{w}$, an imaginary technique that is like sepSim$^{+}$ but additionally supporting $\dot{\Rrightarrow}$ instead of $|\dot{\Rrightarrow}$. In other words, sepSim$^{+}_{w}$ has an even stronger version of the update rule as follows:

SIM-PEEKING-WUPD
$$\frac{{}^{R}\dot{\Rrightarrow} \boxed{\mathrm{R}}\, (\sigma_t, b_t) \overset{*}{\gtrsim} (\sigma_s, b_s)\ \{Q\}}{\boxed{\mathrm{R}}\, (\sigma_t, b_t) \overset{*}{\gtrsim} (\sigma_s, b_s)\ \{Q\}}$$

The concrete instance of resource algebra we use for this example comprises six distinct resources (and their valid sums; omitted in the figure) and they are visualized as follows:

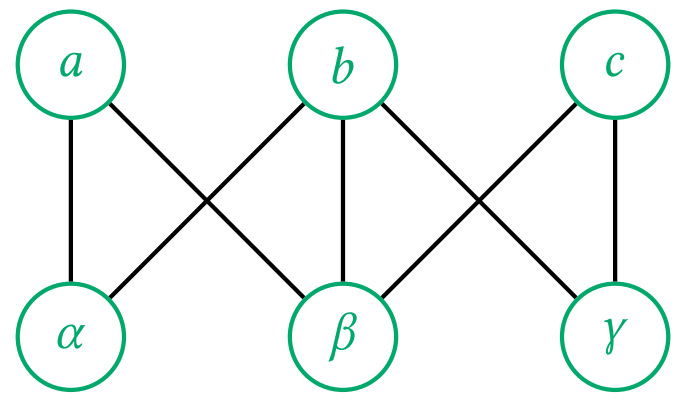

In the example, the target will have exactly one of $\alpha$, $\beta$, and $\gamma$—there cannot be two instances of those—at any given moment. Similarly, the source will have (except for the frozen one) exactly one of $a$, $b$, and $c$. The black lines stand for consistency between them; only the pairs of resources connected with black lines are *valid* (*i.e.*, satisfies $\checkmark(-)$ when summed up), all other cases do not. We are going to prove false by having $a$ and $\gamma$ at the same time (or $c$ and $\alpha$).

Specifically, we consider the following instances:

$$
\begin{array}{ll}
T = (\mathbf{def}\ f() \triangleq \dot{\forall} r \in \mathbb{B}.\ \mathbf{ret}(r)) & M = S = (\mathbf{def}\ f() \triangleq \dot{\exists} r \in \mathbb{B}.\ \mathbf{ret}(r)) \\
\mathrm{cpost}(r) = \mathbf{if}\ r\ \mathbf{then}\ \boxed{\alpha}\ \mathbf{else}\ \boxed{\gamma} & \mathrm{dpost}(r) = \mathbf{if}\ r\ \mathbf{then}\ \boxed{c}\ \mathbf{else}\ \boxed{a} \\
C(f) = (\mathbf{1}, \lambda__.\ \boxed{\beta}, \lambda_.\ \mathrm{cpost}) & D(f) = (\mathbf{1}, \lambda__.\ \boxed{b}, \lambda_.\ \mathrm{dpost}) \qquad E = C \ltimes D
\end{array}
$$

and we will establish $T \sqsubseteq \lfloor M \rfloor_C \sqsubseteq \lfloor S \rfloor_E$.

In establishing first refinement, we crucially use the additional power of $\dot{\Rrightarrow}$: WUPD-UPD-NONDET.

$$T = \quad (\mathbf{def}\ f_t() \triangleq \dot{\forall} r \in \mathbb{B}.\ \mathbf{ret}(r)) \sqsubseteq (\mathbf{def}\ f_m() \triangleq \boxed{\beta}; \dot{\exists} r \in \mathbb{B}.\ \mathrm{cpost}(r); \mathbf{ret}(r)) \quad = \lfloor M \rfloor_C$$

First, we use ASSUME-SRC to get $\boxed{\beta}$. Here, we have $\beta \rightsquigarrow \{\alpha, \gamma\}$: if the given $c$ contains $a$, we pick $\alpha$, and pick $\gamma$ otherwise. Then, using WUPD-UPD-NONDET, we get $\dot{\Rrightarrow} (\exists r \in \mathbb{B}.\ \mathrm{cpost}(r))$. We eliminate this weak update modality (using SIM-PEEKING-WUPD), eliminate the existential quantifier, and witness $r$. After that, we start executing the program part: we execute ANGELIC-TGT, DEMONIC-SRC, and instantiate both cases with $r$. We conclude the proof with ASSERT-SRC and SIM-RET.

In establishing the second refinement, we critically use the additional power of peeking: PEEKING-I.

$$
\begin{array}{rll}
\lfloor M \rfloor_C = & (\mathbf{def}\ f_m() \triangleq \boxed{\beta}; \quad \dot{\exists} r \in \mathbb{B}.\ \mathrm{cpost}(r); \quad \mathbf{ret}(r)) & \\
& \sqsubseteq (\mathbf{def}\ f_s() \triangleq \boxed{\beta} * \boxed{b}; \dot{\exists} r \in \mathbb{B}.\ \mathrm{cpost}(r) * \mathrm{dpost}(r); \mathbf{ret}(r)) & = \lfloor S \rfloor_D
\end{array}
$$

Using $\mathrm{sepSim}_w^+$, proof obligation ($\mathrm{sepSimFun}_w^+$) is proven as follows (we use $\mathcal{I} = \top$ and omit states):

$$\frac{\top}{\forall r.\ (\mathrm{dpost}(r) * \mathrm{cpost}(r)) \vdash (\mathrm{cpost}(r) * \mathrm{dpost}(r))}\ (6)$$

$$\frac{\forall r.\ (\mathrm{dpost}(r) * \mathrm{cpost}(r)) \vdash (\mathrm{cpost}(r) * \mathrm{dpost}(r))}{\forall r.\ \mathrm{dpost}(r) \vdash C; D \Vdash \boxed{\boxed{\beta}}\ \mathbf{ret}(r) \precsim \mathbf{ret}(r)\ \{\mathrm{cpost}(r) \twoheadrightarrow (\mathrm{cpost}(r) * \mathrm{dpost}(r))\}}\ (5)$$

$$\frac{\forall r.\ \mathrm{dpost}(r) \vdash C; D \Vdash \boxed{\boxed{\beta}}\ \mathbf{ret}(r) \precsim \mathbf{ret}(r)\ \{\mathrm{cpost}(r) \twoheadrightarrow (\mathrm{cpost}(r) * \mathrm{dpost}(r))\}}{\forall r.\ ^{\boxed{\beta}} \dot{\Rrightarrow} \mathrm{dpost}(r) \vdash C; D \Vdash \boxed{\boxed{\beta}}\ \mathbf{ret}(r) \precsim \mathbf{ret}(r)\ \{\mathrm{cpost}(r) \twoheadrightarrow (\mathrm{cpost}(r) * \mathrm{dpost}(r))\}}\ (4)$$

$$\frac{\forall r.\ ^{\boxed{\beta}} \dot{\Rrightarrow} \mathrm{dpost}(r) \vdash C; D \Vdash \boxed{\boxed{\beta}}\ \mathbf{ret}(r) \precsim \mathbf{ret}(r)\ \{\mathrm{cpost}(r) \twoheadrightarrow (\mathrm{cpost}(r) * \mathrm{dpost}(r))\}}{\forall r.\ \boxed{b} \vdash C; D \Vdash \boxed{\boxed{\beta}}\ \mathbf{ret}(r) \precsim \mathbf{ret}(r)\ \{\mathrm{cpost}(r) \twoheadrightarrow (\mathrm{cpost}(r) * \mathrm{dpost}(r))\}}\ (3)$$

$$\frac{\forall r.\ \boxed{b} \vdash C; D \Vdash \boxed{\boxed{\beta}}\ \mathbf{ret}(r) \precsim \mathbf{ret}(r)\ \{\mathrm{cpost}(r) \twoheadrightarrow (\mathrm{cpost}(r) * \mathrm{dpost}(r))\}}{\boxed{b} \vdash C; D \Vdash \boxed{\boxed{\beta}}\ \dot{\exists} r \in \mathbb{B}.\ \mathbf{ret}(r) \precsim \dot{\exists} r \in \mathbb{B}.\ \mathbf{ret}(r)\ \{\mathrm{cpost}(r) \twoheadrightarrow (\mathrm{cpost}(r) * \mathrm{dpost}(r))\}}\ (2)$$

$$\frac{\boxed{b} \vdash C; D \Vdash \boxed{\boxed{\beta}}\ \dot{\exists} r \in \mathbb{B}.\ \mathbf{ret}(r) \precsim \dot{\exists} r \in \mathbb{B}.\ \mathbf{ret}(r)\ \{\mathrm{cpost}(r) \twoheadrightarrow (\mathrm{cpost}(r) * \mathrm{dpost}(r))\}}{(\boxed{\beta} * \boxed{b}) \twoheadrightarrow (\boxed{\beta} * C; D \Vdash \boxed{\boxed{\beta}}\ \dot{\exists} r \in \mathbb{B}.\ \mathbf{ret}(r) \precsim \dot{\exists} r \in \mathbb{B}.\ \mathbf{ret}(r)\ \{\mathrm{cpost}(r) \twoheadrightarrow (\mathrm{cpost}(r) * \mathrm{dpost}(r))\})}\ (1)$$

In (1), we cancel $\boxed{\beta}$. In (2), we execute the target and then source with DEMONIC-TGT and DEMONIC-SRC. In (3), the most interesting thing happens: utilizing the fact that we have both $(b \cdot \beta) \rightsquigarrow (a \cdot \beta)$ and $(b \cdot \beta) \rightsquigarrow (c \cdot \beta)$, we apply PEEKING-I and get $^{\boxed{\beta}} \dot{\Rrightarrow} \mathrm{dpost}(r)$, regardless of the value of $r$. In (4), we eliminate the peeking update modality with SIM-PEEKING-WUPD and PEEKING-MONO. In (5), we apply SIM-RET and in (6) the reflexivity of $\vdash$ concludes the proof.

Now, we have established $T \sqsubseteq \lfloor M \rfloor_C \sqsubseteq \lfloor S \rfloor_E$. Take a look at the post-condition of $E$: it is actually equivalent to $\bot$! Regardless of the value of $r$, the postcondition results in a pair of resources that are inconsistent: $(\alpha, c)$ or $(\gamma, a)$. Using this false postcondition, we prove the following false refinement:

$$(\mathbf{def}\ \mathrm{main}_t() \triangleq f(); \mathbf{ret}(0)) \sqsubseteq (\mathbf{def}\ \mathrm{main}_s() \triangleq f(); \mathbf{ret}(1))$$

which in turn implies falsity in the meta-logic (see our Coq development for more detail).

***Discussion.*** This counterexample is basically making some kind of "race condition" between the lower and the upper refinement. In the lower refinement, the application of WUPD-UPD-NONDET illegally (in our view) "peeks" into the source resource. However, this could be very different from which source resource is actually being picked; the source state transition system does not even appear in the lower refinement! Then, its "wrong" view on the source resource is carried through to the upper refinement in the form of program return value and postcondition. However, the actual source resource picked on the upper refinement is opposite from the view of the lower refinement, resulting in a contradiction.

There might be other ways to avoid this counterexample by weakening other aspects of $\mathrm{sepSim}^+_w$. Specifically, removing the peeking update modality and having $\mathrm{sepSim}_w$ would allow us to avoid this problem—only if one can find such a model that does not admit peeking. However, this is non-trivial in CCR and, we believe, any similar wrapper-based approach to conditional refinements.

Suppose that the source precondition is $P * Q$. When we have a target module with precondition $P$, the *operational semantics* of the source should really throw away $P$ in order to disable the peeking update. On the other hand, if we consider another target module with precondition $Q$, the source should throw $Q$ away instead. However, the *operational semantics* of the source is completely *unary*, computed on its own, and is completely independent of what target module we use in refinement later. Without knowing the target module, it is non-trivial to decide which resource to throw away.

To summarize, in our understanding, the peeking update is a rather unwanted and unavoidable side effect of the wrapper-based approach to conditional refinements. It could be possible to remove peeking with a completely different technique from CCR, but that is uncharted territory and outside the scope of this paper. Our choice of strengthening the update modality is sensible in CCR because no proof in CCR relies on WUPD-UPD-NONDET. In Iris, the rule was crucially used for invariant name allocation where the invariants in Iris are much more powerful (supporting impredicativity) than the ones used in CCR. CCR does not support such a feature—doing so would likely require step-indexing—and only supports non-impredicative, static invariants. This obviates the need for the invariant name allocation problem. Also, the usual ghost name allocation can be easily supported using even these simple (non-impredicative and static) invariants.

## C ADDITIONAL PROOF PATTERNS

Fig. 13 shows another example showing the Problem **I**. It does not use separation logic conditions, intendedly, but uses conditions for changing the type of arguments/return values.

The implementation module $T_{\mathrm{BS}}$ implements a bitset using `long` type. The middle abstraction $M_{\mathrm{BS}}$ abstracts it to a natural number and uses operators defined in `Nat` library of Coq. This refinement relies on conditions that relate the natural number with `long`; for instance, consider the condition for `get` in $C$. Using Hoare-Quadruple-like notation, the condition in Fig. 13 is expressed like this:

$$\forall a_m\, a_t.\ \{\ulcorner a_m = l2n(a_t) \urcorner \wedge \ulcorner a_m \in [0, 64) \urcorner\}\ (r_t \leftarrow \mathtt{get}(a_t)) \leq (r_m \leftarrow \mathtt{get}(a_m))\ \{\ulcorner r_m = r_t = () \urcorner\}$$

where $l2n(-)$ converts `long` into $\mathbb{N}$. Since $\mathbb{N}$ has strictly more elements than `long`, conditions like $\ulcorner a_m = l2n(a_t) \urcorner$ is essential, requiring the client to pass only the natural numbers within the range.

Finally, the source module $S_{\mathrm{BS}}$ further abstracts $\mathbb{N}$ into *Flag*, which is a custom algebraic data type defined in Coq. The datatype also has an embedding into $\mathbb{N}$, $f2n(-)$. The conditions for the

$T_{\text{BS}}$

```
def bitset: long = 0

def set(key: long) ≡
  bitset := (bitset & (1 << k))

def get(key: long): long ≡
  ret (bitset | (1 << k))
```

$M_{\text{BS}}$

```
def n: ℕ = 0

def set(k: ℕ): 1 ≡
  n := (Nat.setbit n k)

def get(k: ℕ): 𝔹 ≡
  ret (Nat.testbit n k)
```

$S_{\text{BS}}$

```
private map ∈ Flag → 𝔹 := λk.⊥

def set(k: Flag): 1 ≡
  map := map[k ← ⊤]

def get(k: Flag): 𝔹 ≡
  ret map(k)
```

$C_{\text{BS}}$

$$C(\texttt{set}) \triangleq (\mathbf{1}, \{\lambda a_m\, a_t.\, \ulcorner a_m = l2n(a_t) \urcorner \wedge \ulcorner a_m \in [0, 64) \urcorner\}, \{\lambda r_m\, r_t.\, \ulcorner r_m = r_t = () \urcorner\})$$
$$C(\texttt{get}) \triangleq (\mathbf{1}, \{\lambda a_m\, a_t.\, \ulcorner a_m = l2n(a_t) \urcorner \wedge \ulcorner a_m \in [0, 64) \urcorner\}, \{\lambda r_m\, r_t.\, \ulcorner r_m = l2n(r_t) \urcorner\})$$

$D_{\text{BS}}$

$$D(\texttt{set}) \triangleq (\mathbf{1}, \{\lambda a_s\, a_m.\, \ulcorner f2n(a_s) = a_m \urcorner\}, \quad \{\lambda r_s\, r_m.\, \ulcorner r_s = r_m = () \urcorner\})$$
$$D(\texttt{get}) \triangleq (\mathbf{1}, \{\lambda a_s\, a_m.\, \ulcorner f2n(a_s) = a_m \urcorner\}, \quad \{\lambda r_s\, r_m.\, \ulcorner r_s = r_m \urcorner\})$$

Fig. 13. An implementation $T_{\text{BS}}$, its intermediate abstraction $M_{\text{BS}}$, and its final abstraction $S_{\text{BS}}$ (above). Conditions used in the lower refinement $C$, and conditions used in the upper refinement $D$ (below).

upper refinement, $D$, are largely similar to $C$ except that it relates ℕ and *Flag*. It should be able to further abstract $S_{\text{BS}}$ with separation logic conditions (with points-to connectives giving ownership to specific flags), but for expository purposes, we refrain from doing so.

In this verification, the key benefit of having the middle-level abstraction, $M_{\text{BS}}$, is that it enables *proof reuse* when one comes with another implementation, $T'_{\text{BS}}$. For instance, if $T'_{\text{BS}}$ is using `int32` instead of `long`, the conditions in $C$ should be adjusted accordingly. For the reasons explained in §1, CCR 1.0 does not support proof reuse in this scenario while CCR 2.0 supports it.